\documentclass[onecolumn]{aa} % for a paper on 1 column  
\usepackage[dvipsnames]{xcolor}
\usepackage{graphicx}
\usepackage{txfonts}
\begin{document}

   \title{Diffusion of CH$_4$ in amorphous solid water}

%   \subtitle{I. Overviewing the $\kappa$-mechanism}

   \author{Belén Maté 
          \inst{1}
          \and
       Stephanie Cazaux
       \inst{2,3}
       \and
       Miguel Angel Satorre
       \inst{4}
         \and
       Germán Molpeceres 
       \inst{5}
       \and
       Juan Ortigoso
       \inst{1}
       \and
       Carlos Millán
       \inst{4}
       \and 
       Carmina Santonja
          \inst{4}
          }

   \institute{Instituto de Estructura de la Materia, IEM-CSIC, Calle Serrano 121, 28006 Madrid, Spain \\
              \email{belen.mate@csic.es}
        \and
            Faculty of Aerospace Engineering, Delft University of Technology, Delft, The Netherlands
             \and
            Leiden Observatory, Leiden University, P.O. Box 9513, NL 2300 RA Leiden, The Netherlands\\
            \email{s.m.cazaux@tudelft.nl}
         \and
             Escuela Politécnica Superior de Alcoy, Universitat Polit\`{e}cnica de Val\`{e}ncia, 03801 Alicante, Spain \\
             \email{msatorre@fis.upv.es}
       \and
            Institute for Theoretical Chemistry, University of Stuttgart, 70569 Stuttgart, Germany  \\
             \email{molpeceres@theochem.uni-stuttgart.de}
             }

% \abstract{}{}{}{}{} 
% 5 {} token are mandatory

 \abstract
 {The diffusion of volatile species on amorphous solid water ice affects the chemistry on dust grains in the interstellar medium as well as the trapping of gases enriching planetary atmospheres or present in cometary material.}
   {The aim of the work is to provide diffusion coefficients of CH$_4$ on amorphous solid water (ASW), and to understand how they are affected by the ASW structure.}
   {Ice mixtures of H$_2$O and CH$_4$ were grown in different conditions and the sublimation of CH$_4$ was monitored via infrared spectroscopy or via the  mass loss of a cryogenic quartz crystal microbalance. Diffusion coefficients were obtained from the experimental data assuming the systems obey Fick's law of diffusion. Monte Carlo simulations modeled the different amorphous solid water ice structures   investigated and were used to reproduce and interpret the experimental results.}
   {Diffusion coefficients of methane on amorphous solid water have been measured to be between 10$^{-12}$ and 10$^{-13}$ cm$^2$ s$^{-1}$ for temperatures ranging between 42 K and 60 K. We showed that diffusion can differ by one order of magnitude depending on the morphology of amorphous solid water. The porosity within water ice, and the network created by pore coalescence, enhance the diffusion of species within the pores.The diffusion rates derived experimentally cannot be used in our Monte Carlo simulations to reproduce the measurements.} 
   % conclusions heading (optional)
   {We conclude that Fick's law can be used to describe diffusion at the macroscopic scale, while Monte Carlo simulations describe the microscopic scale where trapping of species in the ices (and their movement) is considered.}


   \maketitle
\section{Introduction}
Icy mantles covering dust grains in dense clouds of the interstellar medium are known to be responsible for the large molecular complexity of our universe. Within those mantles, atoms and molecules can meet and react with higher probability than in the gas phase. The chemical reactivity of interstellar ice is limited by the diffusion of reacting atoms or molecules in water ice, its major component \citep{tielens1982}. In particular the diffusive Langmuir-Hinshelwood mechanism is believed to be the primary formation process on ice mantles (\cite{Irvine2011}). Therefore knowing the barriers to diffusion of the precursors of a reaction is of key importance. However, due to the multiple factors that affect this process, diffusion coefficients are difficult to obtain, both experimentally or theoretically. Due to the lack of accurate information, it is frequently assumed in astrochemical models that the diffusion energy is a fraction of surface binding energy of the molecule \citep{Herbst2001,Garrod2008}. This fraction is very poorly constrained and values between 0.3 and 0.8 are used by the modeling community \citep{hasegawa1992a,cuppen2017,Garrod2013}.

Despite the complexity of the problem, different experiments have been carried out to obtain information about diffusion  \citep{he2018,Ghesquiere2015}. In particular, water ice is one the most investigated systems. Laser-induced Thermal Desorption techniques (LITD) were employed to investigate isotopic diffusion (HDO or H$_2^{18}$0) or molecular diffusion (HCl, NH$_{3}$, CH$_{3}$OH) in and on crystalline water ice (see \citet{Livingston2002} and references therein). More recently, other methodologies have been developed, based on infrared spectroscopy, to study diffusion in amorphous solid water (ASW) (\citet{Mispelaer2013}, \citet{Karssemeijer2014}, \citet{Lauck2015}, \citet{Ghesquiere2015}, \citet{Cooke2018}, \citet{he2018}). Surface diffusion coefficients for CO, NH$_3$, H$_2$CO and HNCO in ASW were given by \citet{Mispelaer2013}. The diffusion of CO in ASW at temperatures between 12 and 50~K was investigated by \citet{Karssemeijer2014} and \citet{Lauck2015}, and information about CO$_{2}$ diffusion was given by \citet{Ghesquiere2015}, \citet{He2017}, and \citet{Cooke2018}. Activated energies of diffusion of CO, O$_2$, N$_2$, CH$_4$ and Ar on ASW were provided in a recent work by \citet{he2018}. In the discussions presented in those papers questions have raised on how the diffusion is affected by factors like amorphous water ice reorganization, porosity, layer thickness, or sublimation processes.

In this work, the experimental methodology proposed by \citet{Mispelaer2013} based on infrared spectroscopy and a new experimental approach based on quartz crystal microbalance (QCMB) measurements, have been combined with Monte Carlo simulations to investigate the diffusion of CH$_4$ in amorphous solid water (ASW).  

%--------------------------------------------------------------------
\section{Experimental part}

  The experiments were carried out in two experimental setups, one at  IEM-CSIC-Madrid and the other at  UPV-Alcoy. The new experimental approach based in QCMB detection was performed at UPV-Alcoy. Similar experiments were conducted in both laboratories to test the viability of the new approach, which has the advantage of allowing measuring diffusion coefficients of non-IR-active species.

\subsection{Madrid experimental setup}

This experimental setup has been described in detail in previous publications (\citet{Mate2018b}, \citet{Galvez2009}, \citet{Herrero2010}). It consists of a high vacuum chamber evacuated to a base pressure of 1 x 10$^{-8}$ mbar and provided with a closed cycle He cryostat. A silicon substrate 1 mm thick is placed in a sample holder in thermal contact with the cold head of the cryostat, and its temperature can be controlled between 15 K and 300~K with 0.5~K accuracy.  Infrared spectra are recorded in normal transmission configuration with a FTIR spectrometer (Bruker vertex70) provided with an MCT detector. In this work we have recorded spectra at 4 cm$^{-1}$ resolution averaging 100 scans. Controlled flows of CH$_4$ (99.95 percent, Air Liquide) and H$_2$O (distilled water, three times freeze-pump-thawed) can be admitted through independent lines to back-fill the chamber. The CH$_4$ line is provided with a mass-flow controller (Alicat) while the H$_2$O gas flow is controlled with a leak valve. Ices were grown by vapor deposition on both sides of the cold Si substrate, at a rate of approximately 1 nm/s.

In order to investigate CH$_4$ diffusion on ASW a procedure inspired by \citet{Mispelaer2013} has been followed. Initially, ices of methane and water were grown at 30 K at rates that range between 1 and 1.5 nm/s. In some cases a two layer system (first CH$_4$, then H$_2$O on top) was generated; in other cases, mixed ices were formed by simultaneous deposition of both gases. Then, the CH$_4$:H$_2$O system was warmed at a controlled rate, between 5 K/min and 20 K/min, to a temperature, T$_{iso}$, above methane sublimation temperature. In most of the experiments performed in this work T$_{iso}$ = 50 K. At this temperature CH$_4$ molecules diffuse through the pores of the amorphous solid water structure until they reach the surface of the ice and sublime instantly. Infrared spectra were recorded as function of time elapsed at T$_{iso}$ (time zero is set when  T$_{iso}$ is reached). The time decay of the intensity of the $\nu_3$ band of CH$_4$ informs of the number of molecules that have moved through the amorphous water ice layer and have left the sample. Different experiments varying ice thicknesses, CH$_4$/H$_2$O ratio, heating rates, and growing configurations (sequential or simultaneous) were performed. A list of them is given in Table 1.

The temperature range chosen to perform the experiments is conditioned by the sublimation temperature of CH$_4$ and experimental limitations. The deposition temperature of 30 K was selected to be the highest one where there is no substantial methane sublimation in our experimental conditions. The isothermal experiment temperature of 50 K was chosen to observe most of the methane diffusion/sublimation in less than one hour. At lower temperatures, CH$_4$ diffusion and sublimation takes longer times, and the growing of a ASW layer due to background water contamination present in the HV setups, will affect the results in a not negligible way. With a base pressure of $\sim$(0.7–1) $\times$ 10$^{-8}$ mbar, water vapor from the chamber deposited on our samples at an approximate rate of $\sim$(3–6) $\times$ 10$^{15}$ molecules cm$^{-2}$ hr$^{-1}$.”

Ice layer thicknesses were determined from the IR absorbance spectra and infrared band strengths. The OH-stretching band at 3200 cm$^{-1}$ and the $\nu_3$ mode at 1300 cm$^{-1}$ were chosen for H$_2$O and CH$_4$, respectively. The absorption coefficients of those bands are $A_{3200}$ = 2.0 $ \times 10^{-16}$ cm molec$^{-1}$ (\citet{Mastrapa2009}), and $A_{1300} = 6.5 \times 10^{-18}$ cm molec$^{-1}$ (\citet{Molpeceres2017}), respectively. Densities of 0.65 g cm$^{-3}$ (\citet{Dohnalek2003}) for ASW and 0.46 g cm$^{-3}$ for CH$_4$ (\citet{Molpeceres2017}) were assumed, for ices grown by background deposition at 30 K. In the case of ice mixtures grown by codeposition, the determination of ASW film thickness is less accurate, due to the lack of information about band strengths and densities for these mixed systems. However, considering the pure species values for the band strengths is expected to lead to less than 20 percentage error (\citet{Kerkhof1999}, \citet{Galvez2009}). For a given molecular ratio, we have taken the corresponding average density. The estimate uncertainty in the codeposited mixtures ice thickness is 20\%.

\begin{table*}[htbp]
\begin{center}
\begin{tabular}{|c|c|c|c|c|c|c|c|}
%{|l|l|l|}
\hline
Exp. &   & $L_{ASW}/L_{CH_4}$ & ramp & T$_{iso}$ &  \% CH$_{4}$ &  \% CH$_{4}$ &  \% CH$_{4}$ \\
 &   & nm/nm & K/min & K &  30 K & ini T$_{iso}$ & fin T$_{iso}$\\
\hline \hline

M1	& s	 & 146/19 & 5 & 50		& 17.3 &	10.6 &	7.3 \\ \hline
M2	& s	 & 228/17 & 5	 & 50    & 6.5	& 5.9 &	3.1 \\ \hline 
M3	& s	 & 275/36 & 20 & 50	& 13.4	& 10.5 & 8.2 \\ \hline
M4	& s	 & 344/16 & 10 & 50	& 3.6	& 3.6 &	2.4 \\ \hline
M5	& s	 & 410/40 & 10 & 50	& 8.9	& 7.7 &	5.3 \\ \hline 
M6	& s	 & 453/47 & 5 & 50		& 9.2	& 8.3 &	5.8 \\ \hline

M7	& c	 & 185	 & 10 & 50	& 13.8 &	12.1 &	7.3 \\ \hline
M8	& c	 & 201	 & 5 & 50		& 13.2 &	11.4 &	7.0 \\  \hline
M9	& c	 & 212 & 5 & 50		& 12.0 &	8.8 &	5.7 \\ \hline

M10	& c	 & 567     & 5 & 55		& 7.6  &	6.3  & 4.6 \\ \hline
M11	& s	 & 487/37 & 5 & 55		& 6.5   & 	6.0	 &  4.9 \\ \hline
M12	& c	 & 467     & 5 & 60	& 5.3  &	4.8  & 3.9 \\ \hline

 \end{tabular}
\caption{List of experiments performed at IEM-CSIC-Madrid. s or c stands for sequential or codeposited experiments, respectively. $L_{CH_4}$ and $L_{ASW}$ refer to thickness of CH$_4$ and H$_2$O layers (s experiments), or to the mixture ice thickness (c experiments), at 30 K. Columns six, seven and eight present the number of molecules ratio in percentage (100x$N_{CH_{4}}$/$N_{H_{2}O}$) of CH$_{4}$ at 30 K, at the beginning, and at the end of the isothermal experiment.}
\label{table:1}
\end{center}
\end{table*}

\subsection{UPV-Alcoy experimental setup}

The experimental setup has been described in detail in previous publications (\citet{Luna2012}). It consists of a high vacuum chamber that reaches $5 \times 10^{-8}$ mbar background pressure, where a closed He cryostat allows cooling down a sample holder to 13~K. A quartz crystal microbalance (QCMB) is located in thermal contact with the cold head, acting as part of the sample holder. A resistor permits to vary the temperature of the cold head between 13~K and room temperature, within 0.2~K by means of a ITC-301 temperature controller (Oxford Instruments) and two silicon diodes (Scientific Instruments). One of them is located just below a QCMB and the other just at the end of the second stage of the cold finger of the Leybold Cryostat. The system is provided also with an He-Ne double laser interference system ($\lambda$ = 632.8 nm), that enters the chamber through KBr windows to impinge on the QCMB quartz surface. By measuring the laser interference patterns during ice growing, it is possible to determine the real part of the refractive index of the ices deposited, at the laser wavelength, and the thicknesses of the ice layers generated. 
When vapors are introduced in the vacuum chamber an ice layer grows on top of the QCMB. The QCMB measures any variation in mass deposited on the surface of the quartz crystal. As the mass increases (adsorption, deposition) the vibration frequency of the quartz decreases; if the mass decreases (desorption) the quartz vibrational frequency increases. There is a linear dependency between $ \Delta F $, the variation of the quartz crystal frequency, and $ \Delta m$, the mass variation on the quartz crystal surface, the so-called Sauerbrey relationship:  
%Sauerbrey's Eq
\begin{equation}
\Delta F = -S\Delta m
\label{eq:1}
\end{equation}
where $S$ is the Sauerbrey's constant.

The procedure employed to investigate CH$_4$ diffusion is in essence the same than the one used in the Madrid experiments. Ices of CH$_4$ and H$_2$O were grown by codeposition or sequential deposition (first CH$_4$, then H$_2$O on top) at 30 K, a temperature well below CH$_4$ sublimation, and then they were warmed up to T$_{iso}$, above methane sublimation temperature. At this T$_{iso}$, the amount of methane that diffuses through the porous structure of ASW and leaves the ice is monitored registering the frequency variation of the QCMB. As explained above, the variation of the quartz crystal frequency is proportional to the variation of the mass deposited on its surface (see equation \ref{eq:1}). In particular, a mass loss along time will be observed as a QCMB frequency increases versus time. In this case, instead of recording the time evolution of the IR spectra of the ice, the QCMB is providing this information. It is interesting to notice that this new experimental approach, opens the possibility to study diffusion of volatile species that do not have IR spectrum, like H$_2$, N$_2$ or O$_2$.

Special care was made in using similar deposition rates, CH$_4$:H$_2$O ratios, and heating ramps in the experiments performed in both laboratories, in order to facilitate comparison of the results. Nonetheless, larger ASW layer thickness, up to 1 micron, were covered in the experiments performed in Alcoy. This set of Alcoy experiments has been named A1 to A10, and are listed in Table 2.

In an attempt to study the influence of the temperature on the diffusion coefficient, and extract the energy of the diffusion barrier and the preexponential factor, avoiding the effect of having distinct water ice structures, a different set of experiments was designed. In this case, ASW was grown at 50 K (background deposition), was kept several minutes at that temperature, cooled down to 30 K to deposit a CH$_4$ layer on top, and finally warmed to the desired T$_{iso}$, that ranged between 42.5 K and 52.5 K. At T$_{iso}$ the mass loss versus time is monitored with the QCMB.  Since ASW ice grown at 50 K do not change its structure for being cooled to 30 K and then warmed back to 50 K, in these experiments CH$_4$ diffusion is measured  in ASW ices of identical structure. When CH$_4$ is deposited on ASW at 30 K, it diffuses through ASW pores. During the warm up process to  T$_{iso}$, most of the methane is sublimated and only the fraction that is within the pores is retained. Since the amount of CH$_4$ retained is very low, the measurements have larger errors than with the previous methodology. These experiments were labeled AA1-AA5.

\section{Experimental results}

In this section, a first subsection is dedicated to discuss the morphology of the different ices investigated via the analysis of their OH-dangling-bond IR spectra. This analysis gives some qualitative information about the quality of the approximations made for the determination of the diffusion coefficient in the following subsections. In a second subsection, the isothermal experiments are presented, and the last one is devoted to describe the models employed to extract diffusion coefficients from them.

\subsection{ASW morphology and temperature reorganization}

Porosity of amorphous solid water formed by vapour deposition on a cold surface depends on the growing conditions. It is affected by the surface temperature, the vapour pressure, or the angle at which the molecules hit the surface (\citet{Berland1995}, \citet{Brown1995}, \citet{Dohnalek2003}, \citet{Bossa2014}). The intrinsic density of ASW also changes, it varies between 1.1 g cm$^{-3}$ for ices grown below 35 K and 0.94 g cm$^{-3}$ for ices grown at T > 70 K. High density amorphous ice evolves to low density ice between 38 K and 68 K, (\citet{Narten1976a}, \citet{Jenniskens1995}).  
The measured average density of ASW grown by background deposition at 30 K is 0.65 g cm$^{-3}$ \citet{Dohnalek2003}. Assuming its intrinsic density is that of the low density ice (0.94 g cm$^{-3}$), its porosity can be estimated via the expression: $\rho= 1-(\rho_a/\rho_i)$= 0.31.The porous structure formed at this temperature has also been investigated and found to be mostly microporous \citep{Raut2007,cazaux2015,he2019}. 

In most of the experiments performed in this work, ices grown at 30 K were warmed to 50 K. During the heating process, a reorganization of the water molecules takes place. Pore clustering \citep{Raut2007,cazaux2015,he2019}, combined with a reorganization of the intrinsic structure of water molecules, is expected to happen. As a result the average density of the ice increases \citep{Berland1995,Dohnalek2003}. Moreover, the presence of CH$_4$ in codeposited CH$_4$:H$_2$O mixtures, will affect ASW morphology at 30 K, and upon warming. These effects will be investigated in this work by Monte Carlo simulations (see section 6). 

\begin{figure}
    \centering
    \includegraphics[width=8cm]{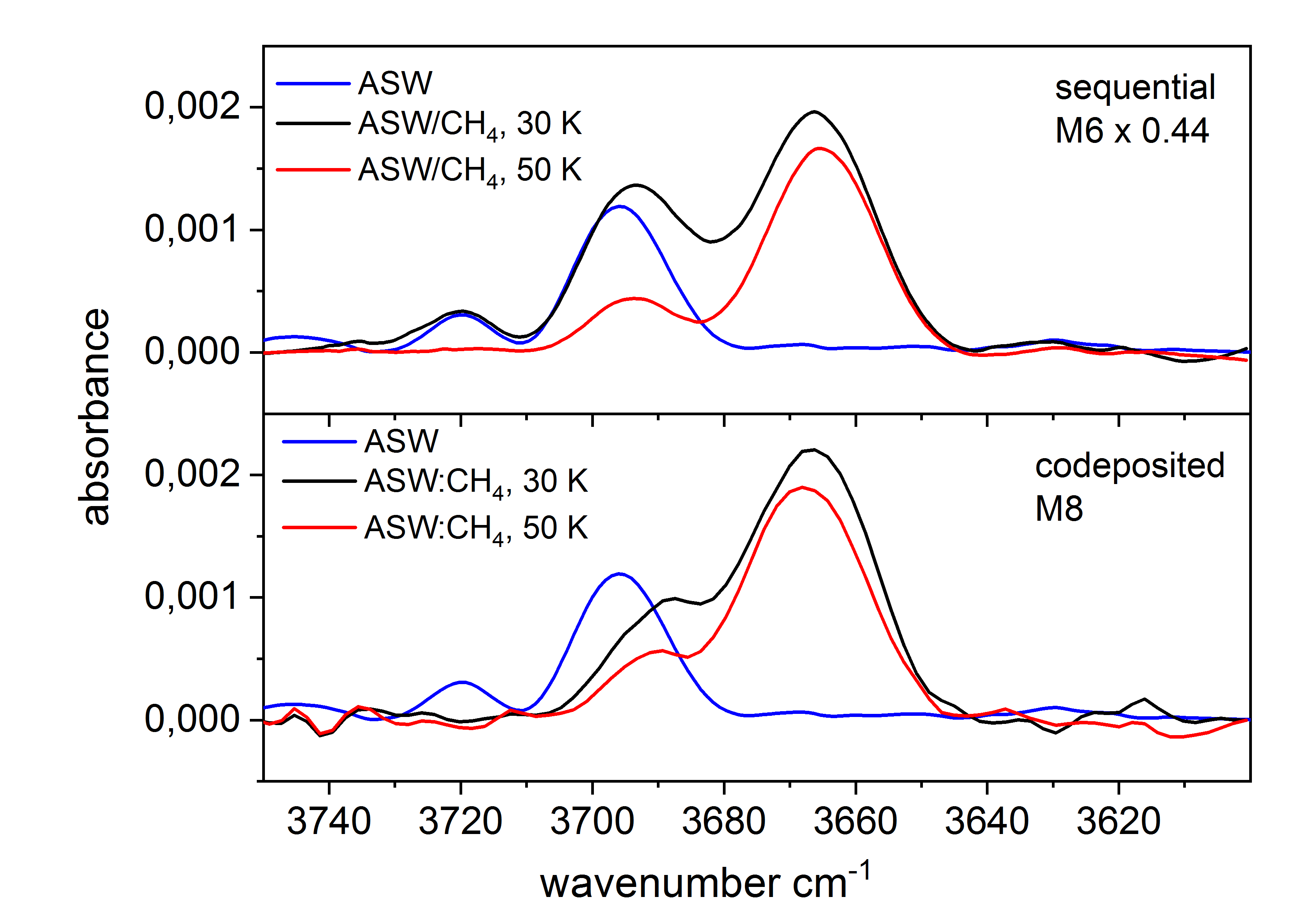}
    \caption{IR spectra of the dangling bond vibrations of water for CH$_4$:H$_2$O mixtures grown at 30 K and warmed to 50 K at 5 K/min, compared with the spectrum of pure water ice grown at 30 K. Top panel: layered ice:H$_2$O on top of ASW. Bottom panel: codeposited ice. All spectra correspond to ASW layers of about 200 nm}
    \label{fig:1}      
\end{figure}

Free OH bonds, know as OH-dangling bonds (DB),  appear on the surface of the porous of the amorphous water ice. They provide information about the porosity of the ice, and about how CH$_4$ molecules will diffuse through the porous structure. Figure \ref{fig:1} shows the DB spectral region for sequential and codeposited H$_2$O:CH$_4$ ices grown at 30 K and warmed to 50 K. On one hand, in sequential experiments it is observed that even at 30 K CH$_4$ molecules diffuse through the ASW structure, as was previously reported  \citep{Raut2007,Herrero2010}. In particular, a new peak appears in the water DB region (see top panel, Figure \ref{fig:1}). The pure water DB double peak at 3720 cm$^{-1}$ and 3696 cm$^{-1}$ (blue trace), transforms into a triple peak feature with maxima at 3720 cm$^{-1}$, 3692 and 3665 cm$^{-1}$ (black trace). An overall increase in DB peaks intensity is also noticed. Since the number of water DBs has not changed, the intensity increase must be due to an increase of the infrared band strength of the dangling OH mode when it is interacting with CH$_4$.
This effect was already shown in previous publications (\citet{Herrero2010}). The peak at 3665 cm$^{-1}$ is characteristic of the interaction of CH$_4$ molecules with dangling bonds of water. Its presence evidences that CH$_4$ molecules have already diffused into ASW pores at 30 K. Subsequently, during the warm up of the ice to 50 K, the pure ASW 3720 cm$^{-1}$ peak disappears, and an intensity decrease of the DB features is noticed (red trace in top panel Figure \ref{fig:1}). The 3665 cm$^{-1}$ peak strongly dominates the profile at 50 K. The intensity decrease reflects that some CH$_4$ is lost upon warming, but also it is related to a decrease in the specific surface area (SSA) of the ASW ice when warmed from 30 K to 50 K, as shown by simulations (see section 6).

On the other hand, the lower panel in Figure \ref{fig:1} shows the DB features observed when a CH$_4$:H$_2$O ice is grown by simultaneous deposition at 30 K (black trace). In particular, the pure ASW DB feature at 3720 cm$^{-1}$ and most of the 3696 cm$^{-1}$ peak (blue trace) are missing in the 30 K codeposited ice. This indicates that methane molecules are already covering most of the ASW pore surface in this kind of mixtures. When the mixture is warmed to 50 K, no changes in the DB spectral profile occur, and only a decrease in intensity is observed (red trace). This decrease, similarly to what happened in sequential ices, is due to some CH$_4$ loss, but also reflects a decrease of the ASW specific surface area. In relation with the overall intensity of the DB features in CH$_4$:H$_2$O codeposited ices, it should be pointed out that, as described in previous works (\citet{Galvez2009}, \citet{Herrero2010}), the DB intensity in the mixtures is always larger than that of the pure ice, and increases with CH$_4$ concentration.

\begin{figure}
    \centering
    \includegraphics[width=8cm]{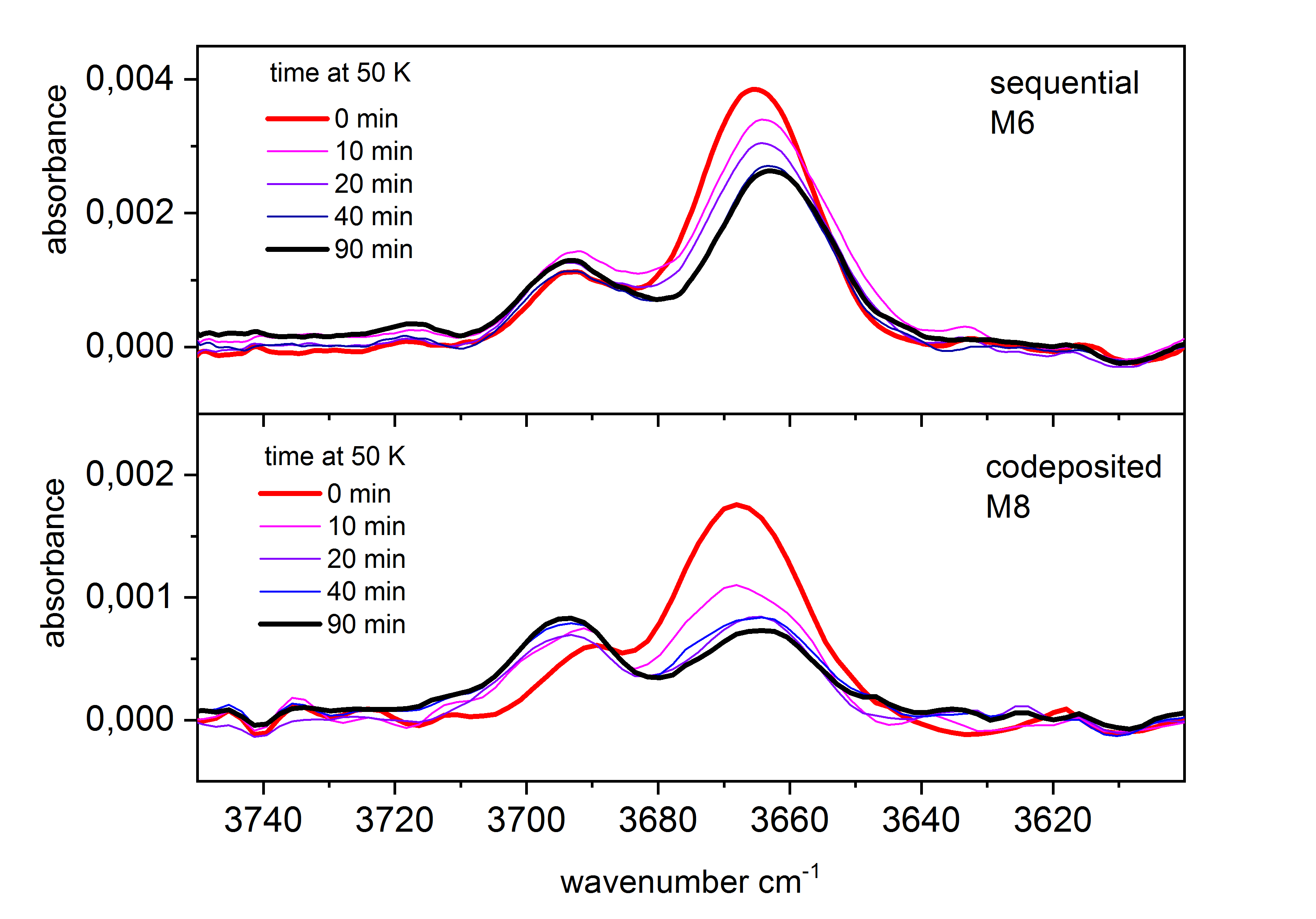}
    \caption{IR spectra of the dangling bond vibrations of water for CH$_4$:H$_2$O mixtures grown at 30 K and warmed to 50 K. Spectra taken at 50 K, at different times during the isothermal experiments. Top panel: layered ice:H$_2$O on top of ASW. Bottom panel: codeposited ice. }
    \label{fig:2}      
\end{figure}

Most of the water ice reorganization takes place during the warm up process from 30 K to T$_{iso}$, nonetheless ASW morphology changes may continue during the time elapsed at that T$_{iso}$. This is a something to be aware with when analyzing the experimental data, since it will be assumed (see section on diffusion modeling) that methane molecules move through the pore surface of an ASW fix structure. The DB evolution at T$_{iso}$ might give an idea of the importance of that reorganization. Figure \ref{fig:2} shows the DB feature for sequential and codeposited ices at different times during the isothermal experiment. The main spectral variation shown in both panels of Figure \ref{fig:2} is a decrease of the 3665 cm$^{-1}$ band, that is associated with  CH$_4$ loss. The rest of the DB feature suffers minor modifications indicating ASW reorganization is not significant during the isothermal experiment. 

 Following the intensity decay of the 1300 cm$^{-1}$ band, it was possible to quantify the fraction of CH$_4$ that leaves the ice during the heating of the ice mixture via IR spectroscopy. Table 1 displays the CH$_4$/H$_2$O number of molecules ratio at 30 K; at T$_{iso}$ at the beginning of isothermal experiment; and at the end of the isothermal experiment. Looking at columns 6 and 7 it is observed that a fraction CH$_4$ is already lost at the beginning of T$_{iso}$, which indicates CH$_4$ is distributed through the whole ASW top layer at this time. Comparing columns 7 and 8 it can be seen that the fraction of methane that left the ice during the isothermal experiment is small, a 20 \% on average. Most methane remains trapped in the ASW structure at the end of the isothermal experiment, both in sequential and in codeposited ices, as the experiments monitor only the methane that diffuse through the open canals (pores) of this structure. As it is well known from thermal programmed desorption measurements of H$_2$O:CH$_4$ mixtures, a release of methane around 40-50 K is followed by a second desorption peak around 140 K  (associated to the amorphous-to-crystalline water phase change, "volcano" desorption), and a last desorption peak appears together with water sublimation, revealing that even a small fraction of CH$_4$  stays in the water ice structure until its sublimation (see for example \citet{May2013}).

\subsection{Diffusion of CH$_4$ in ASW.}

\subsubsection{Madrid experiments}

\begin{figure*}[htbp]
 \centering
\includegraphics[width=18cm]{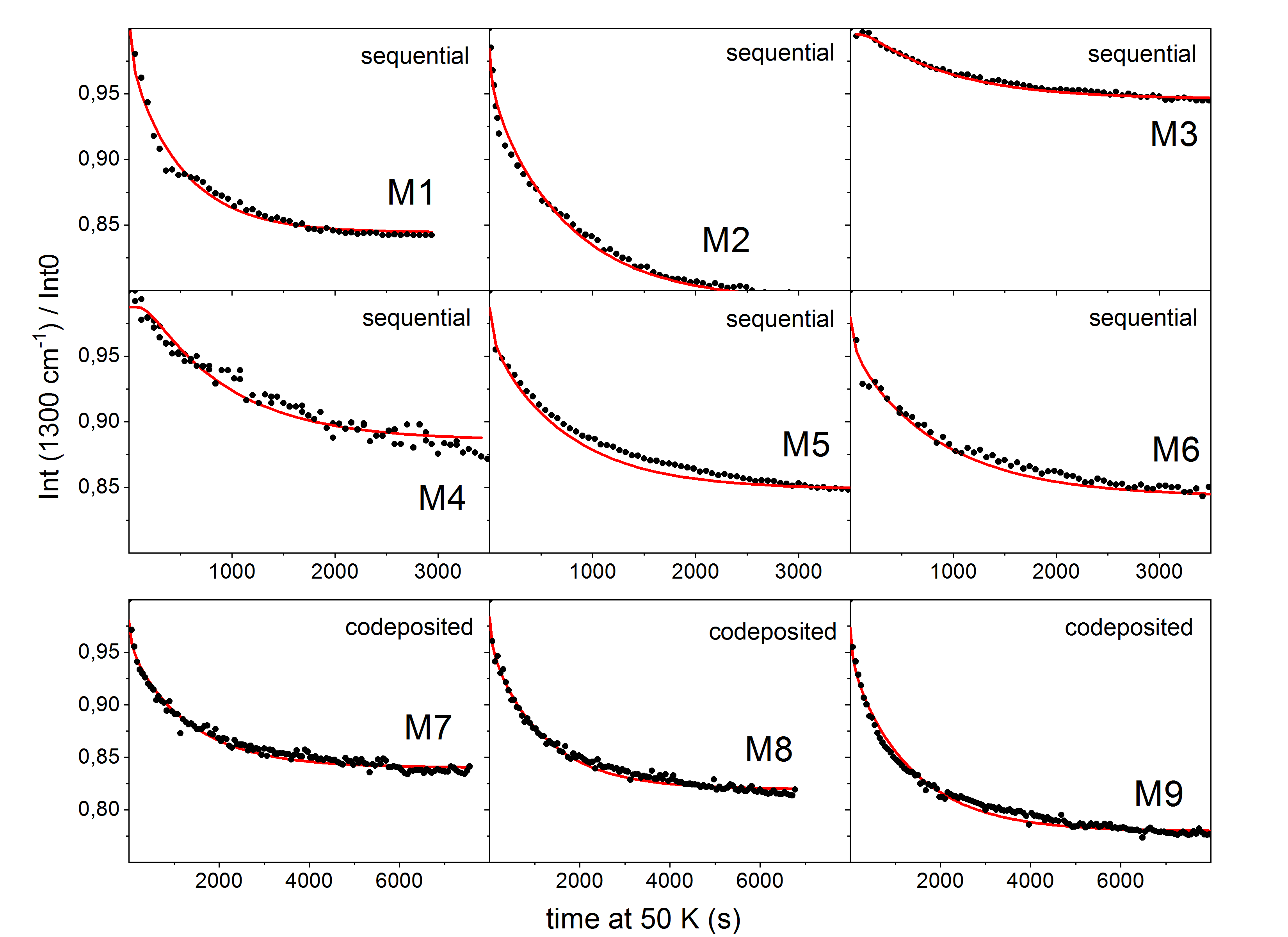}
\caption{Normalized decay of the intensity of the 1300 cm$^{-1}$ band of CH$_4$ versus time elapsed at 50 K. Diamonds: experimental data, line:fit to the second's Fick  diffusion law.}
\label{fig:3}      
\end{figure*}

\begin{figure}
 \centering
\includegraphics[width=10cm]{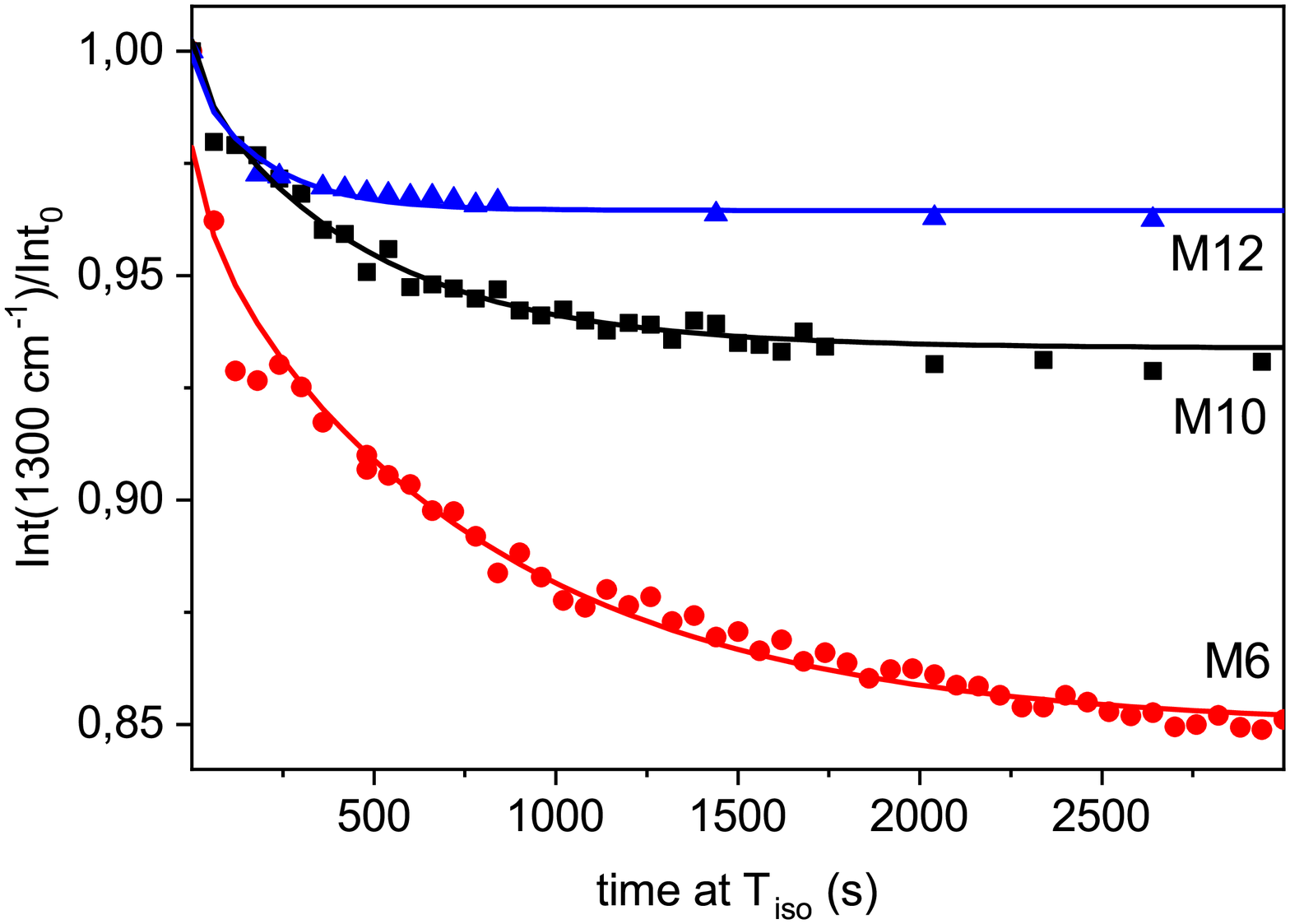}
\caption{Normalized decay of the intensity of the 1300 cm$^{-1}$ band of CH$_4$ versus time elapsed at 50 K, 55 K and 60 K. Scatter points: experimental data, lines: fit to the second's Fick diffusion law.}
\label{fig:4}      
\end{figure}

Figure \ref{fig:3} shows the decay of the intensity of the 1300 cm$^{-1}$ band of methane versus time elapsed at 50 K for the set of experiments presented in Table 1. The integral has been normalized to the band intensity at the beginning of the isothermal experiment. The  panels in the first two rows correspond to sequential deposited ices, with a water ice thickness layer that varies between 145 nm and 480 nm. Panels in the bottom row display the codeposited ice experiments with ASW thickness of $\sim$ 200 nm. 
Three isothermal experiments, performed at 50 K, 55 K and 60 K, corresponding to layers $\sim$ 500 nm thick, are presented in Figure \ref{fig:4}. 

In order to extract diffusion coefficients from these data we have model the decays presented in Figures \ref{fig:3} and \ref{fig:4} using Fick's second law of diffusion, as will be described below in subsection 3.3.

\subsubsection{Alcoy experiments}

Figure 5 presents the variation of the frequency of the QCMB, showing the signal increase, which is proportional to mass decrease, versus time elapsed at 50 K. In an experiment, the larger the frequency variation, the larger the fraction of methane molecules that have been sublimated from the ice (Equation \ref{eq:1}). Frequency variations have been scaled to plot these experiments in the same graph. Experiments cover the range of thicknesses studied, between 300 nm and 1000 nm. These experiments have been analyzed using the first Fick's law.

\begin{figure}
 \centering
\includegraphics[width=10cm]{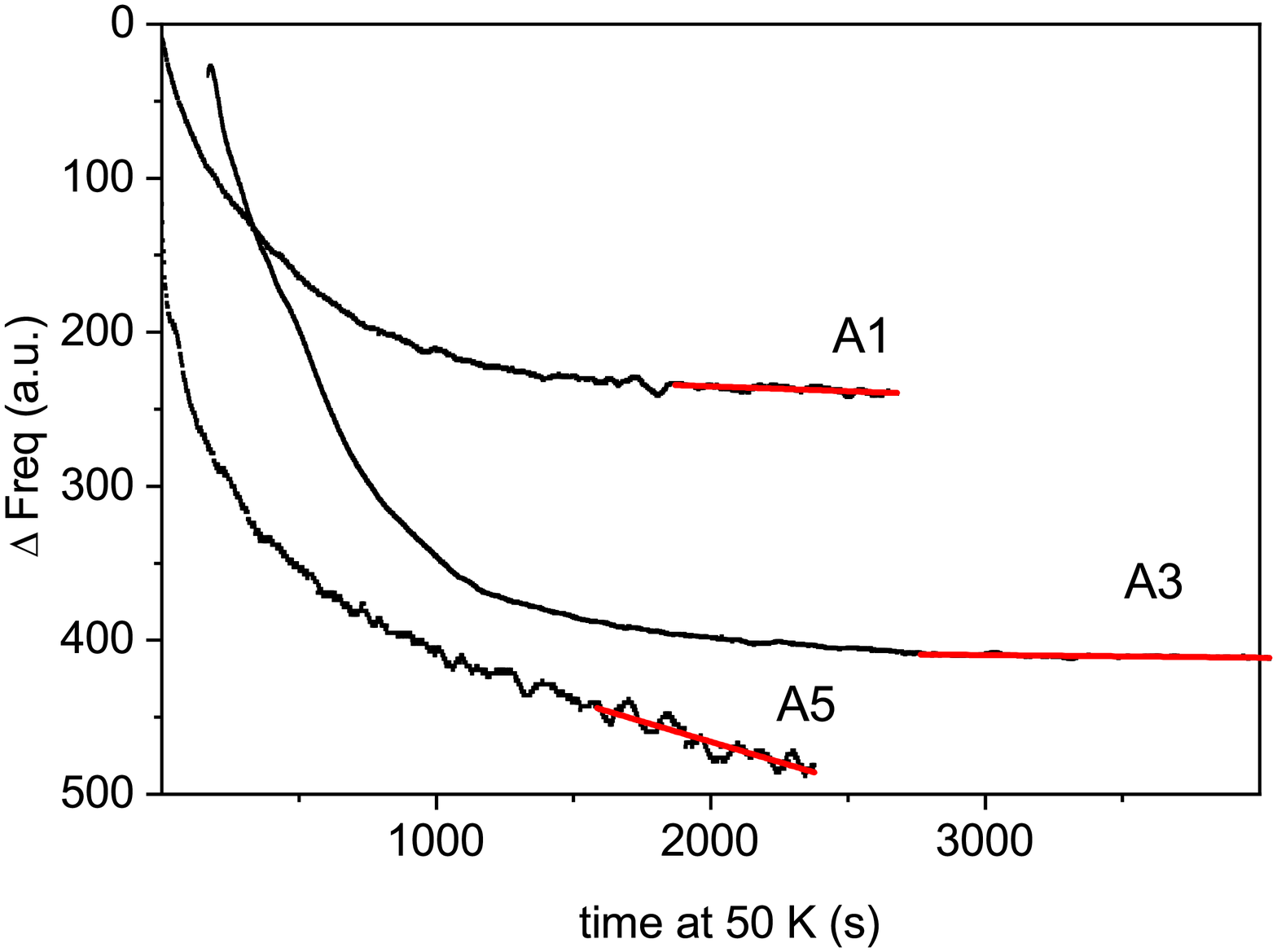}
\caption{Diffusion of methane from ASW ice grown at 30 K. QCMB signal variation versus time elapsed at 50 K. Black lines: experimental data, Red lines:linear fit employed to derive diffusion coefficients form the first Fick's law}
\label{fig:5}      
\end{figure}

\subsection{Fickian diffusion modeling}

\subsubsection{Fick's second law of diffusion}

Fick’s second law of diffusion, that relates the unsteady diffusive flux to concentration gradient, has been applied to obtain the diffusion coefficient, $D(T)$, at a given temperature T. In a one-dimensional system where $D(T)$ does not depend on $z$, i.e., ${\partial D(T)/ \partial z = 0}$, it is given by:
\begin{equation}
\frac{\partial n(z,t)}{\partial t} = D(T)\frac{\partial ^2{n(z,t)}}{\partial z^2}
\label{eq:3}
\end{equation}
where $n(z,t)$ is the concentration of the diffusing species at depth $z$, in the ASW ice at a given time $t$. Equation \ref{eq:3} is used to describe the methane molecules diffusion along a fixed ASW layer of known thickness, i.e., in a direction perpendicular to the surface substrate. This approach was shown to give good results to describe similar systems by other authors \citep{Mispelaer2013, Karssemeijer2014, Lauck2015, he2018}. Some initial conditions must be imposed in order to obtain adequate solutions of this equation. It will be assumed that all CH$_4$ molecules that reach the surface of the ice layer desorb immediately, therefore $n(L,t)=0$, where L is the ASW ice thickness. Since no CH$_4$ molecules can escape from the bottom of the film, ${\partial n(0,t)/ \partial t = 0}$. As a last condition, the concentration of CH$_4$ will be assumed homogeneous within the amorphous solid water ice layer at the beginning of the isothermal experiment, $n(z,0)=n_0$. In particular, for sequential experiments, this last condition implies that during the heating process methane has diffused into the ASW top layer filling the pores homogeneously. Sequential experiments were modelled also imposing a different boundary condition. It was assumed that methane is no homogeneously distributed in the ASW layer, but that it is in a bottom layer with the ASW layer on top. The fits obtained in this way were worse than those found with the previous approximation. The solutions of equation \ref{eq:3} for the constraints of our experiment are (\citet{crank1975}):
\begin{equation}
n(z,t) = \sum_{n=1}^{\infty}{\frac{2n_0(-1)^i}{\mu_i\, L}}\cos{(\mu_i\,z)\exp\left({-\mu_{i}^{2}\,D\,t}\right)}
\label{eq:4}
 \end{equation}
where
\begin{equation}
\mu_i = \frac{(2i+1)\pi}{2L}.
\end{equation}
From these expressions, the column density of methane molecules in the ice can be obtained by integrating the concentration $c(z,t)$ over the ice thickness $L$. The column density is proportional to
the area of a particular CH$_4$ band in the absorbance spectra, $A(t)$, and therefore, integrating over $z$, the solution can be expressed as (\citet{Karssemeijer2014}):
\begin{equation}
A(t) = s+\sum_{n=1}^{\infty}{\frac{2(A_0-s)}{\mu_i^2\,L^2} \exp\left({-\mu_{i}^{2}\,D\,t}\right)}
\label{eq:5}
\end{equation}
where $A_0$ is the initial band area and $s$ is an offset related with the amount of methane that remains trapped in the ice.

Unweighted least squares fitting of the experimental data presented in Figures 3 and 4 to equation \ref{eq:5} have been performed. 
The solutions are represented with solid lines in the mentioned Figures. 
The diffusion coefficients obtained are given in Table \ref{table:2}. 

\subsubsection{Fick's first law of diffusion}
Fick's first law describes the relation between the flux of diffusion through a surface and the concentration gradient perpendicular to that surface, under steady-state conditions. Fick's first law is given by:
\begin{equation}
J = -D(T) \frac{dn(z)}{dz}   
\label{eq:6}
\end{equation}
were $J$ is the flux in molec cm$^{-2}$ s$^{-1}$, and $n(z)$ is the gas concentration in molec cm$^{-3}$, and both magnitudes do not change with time.

Isothermal experiments provide the  number of methane molecules, $n(t)$, that leave the ice at a particular time. Therefore, it is possible to calculate the methane flow, $J(t)= dn(t)/dt$. 
Looking at the final steps of the diffusion experiments, the variation of the flux with time is very slow, and it can be considered almost constant. 
Then, in that region Fick's first law can be applied to extract  $D(T)$. Another approximation has to be made in order to estimate the concentration gradient. At times where the methane flow is considered constant, a constant linear distribution of methane through the ASW layer thickness will be assumed. The CH$_4$ concentration will be expressed as: $n(z)= a + bz$,
 where $n(0)=a$ at the cold deposition surface, and $n(L)=0$ at the ice surface. Consequently, the column density of molecules, $N$, (molec cm$^{-2}$) in the ice will be given by: 
\begin{equation}
N= \int \limits_{z=0}^{z=a}{(a + bz) dz} 
\label{eq:7}
\end{equation}
From this expression and Fick's first law, the diffusion coefficient is given by:
\begin{equation}
D(T)= -J \frac{L}{2N}
\label{eq:8}
\end{equation}
where $L$ is the ASW layer thickness and $J$ the methane flow. There is always a fraction of methane molecules that remain trapped in the ice at the end of the diffusion experiments. 
For the estimation of $N$, that fraction is not taken into account, and only the CH$_4$ that intervenes in diffusion is considered. 
To obtain the diffusion coefficients given in Table \ref{table:2} for the A1-A10  and AA1-AA5 experiments, a linear fit of the QCMB signal at the final part of the isothermal experiments has been performed (see Figure \ref{fig:5}). The slope of the fit gives the methane flow that is necessary to derive $D(T)$ with equation \ref{eq:8}. 

\begin{table*}[htbp]
    \begin{center}
    \begin{tabular}{|c|c|c|c|c|}
    \hline
    Experiment &  $T_{iso}$  & $L_{ASW}/L_{CH_4}$ & D(2$^{nd}$ Fick)  & D(1$^{rst}$ Fick)   \\
     & (K) & (nm/nm) & (10$^{-13}$ cm$^2$ s$^{-1}$) & (10$^{-13}$ cm$^2$ s$^{-1}$)  \\  \hline \hline
     
    M1	  & 50	 & 146/19		& 1.7 & 1.0	\\ \hline      
    M2	  & 50	& 228/17		& 2.7 & 1.3	  \\ \hline    
    M3	  & 50	 & 275/36		& 3.7 & 1.6 	    \\ \hline   
    M4	  & 50	 & 344/16	& 2.5	& 2.1 \\ \hline   
    M5	  & 50	 & 410/40		& 8.1 & 4.0 	  \\ \hline     
    M6	  & 50	& 453/47	& 10.3 & 4.0 	 \\ \hline      
    		  & 		& 		 &     &      \\ \hline            
    M7	  & 50	 & 185	    	& 1.4 & 0.9 	     \\ \hline
    M8	  & 50	 & 201	    	& 1.8	& 0.6    \\ \hline  
    M9	  & 50	 & 212	    	& 1.8 & 0.8  \\ \hline     
    		  & 	      &     &        &        \\ \hline         
    M10 &   55	 &  567		& 26.9 &     \\ \hline  
    M11 &   55	 &  487/37	& 15.4 &   \\ \hline 
    M12 &   60   &  469       & 44.9  & 	  \\  \hline
         &      &            & &   \\ \hline \hline
    A1	&	50  & 333 & & 9   \\ \hline
    A2 	&	50  & 333 & & 1.2  \\ \hline
    A3 	&	50  & 476  & & 6.96 	 \\ \hline
    A4 	&	50 & 520 & & 6 	 \\ \hline
    A5 	&	50  & 827 & & 30  	 \\ \hline
    A6 	&	50  & 913 & & 39 \\ \hline
    A7 	&	50  & 1026 & & 57  	 \\ \hline
    A8 	&	50  & 1033 & & 42 \\ \hline
    A9 	&	50  & 223/169 & & 1.56	  \\ \hline
    A10 	&	50  & 378/33  & & 5.38 	 \\ \hline
         &      &     $L_{CH_4}/L_{ASW}$        & &   \\ \hline \hline
    AA1  &	42.5  & 335/951   & & 0.226   \\ \hline
    AA2  &	45    & 355/946   & & 1.03   \\ \hline
AA3  &	47.5  & 353/930   &  & 0.929   \\ \hline
AA4  &	50    & 358/945   & & 3.62  \\ \hline
AA5  &	52.5  & 356/953   & & 4.23   \\ \hline
 
\end{tabular}
\caption{CH$_4$ diffusion coefficients on ASW grown at 30K (M1-M12 and A1-A10) or at 50 K (AA1-AA5), obtained with two different approximations, Fick's second law or Fick's first law, as indicated. When only one thickness appear in column three it correspond to the ASW ice thickness in codeposited ices, estimated as if no CH$_4$ were present.}
\label{table:2}
\end{center}
\end{table*}

\section{Discussion of experimental results}

Inspecting the results presented in Figures 3, 4 and 5, and in Table 2, several conclusions can be extracted.

1) First and second Fick's law diffusion coefficients.
Both approximations, even though first Fick's law can only be applied to the final part of the isothermal experiments and imposes stronger restrictions, give comparable diffusion coefficients. When both approximations are applied to the same set of experiments (M1-M9), the diffusion coefficients found with the first law are slightly smaller than those found with the second law (see Table \ref{table:2}). The new methodology proposed, based on QCMB detection (exp A1-A10), gives results consistent with those obtained from IR spectrosocopy (M1-M12).

2) CH$_4$ diffusion coefficient dependence on ASW ice morphology.
The heating ramp  will affect ASW morphology at T$_{iso}$. However, from the inspection of the experiments presented in Table 1, where the  ramp was varied between 5 K/min and and 20 K/min, it was not possible to extract any conclusion about its effect on methane diffusion.  It looks like the morphology variations caused by the different heating ramps are not significant enough to be manifested in the $D(T)$ coefficient.

From inspection of Figure \ref{fig:3}, codeposited experiments seem to be better described by the second Fick's law than sequential experiments. Although all fits in Figure \ref{fig:3} are satisfactory, in order to get proper fits of sequential experiments, only times up to 3000 seconds were considered. However, codeposited experiments could be properly fitted in all the experimental time interval (up to 8000 s). This behavior is related to differences in the morphology of the ice. Monte Carlo simulations have been performed for both ASW structures, pure and codeposited ASW, shown in Figure \ref{fig:9} (left panels) and Figure \ref{fig:10}, respectively, illustrate those changes. Diffusion coefficient obtained for codeposited experiments (M6-M9) vary between 1.4 and 1.8 10$^{-13}$ cm$^2$ s$^{-1}$, and those for sequential experiments with similar layer thickness (M2 and M3) are slightly larger, varying between 2.7 and 3.7 10$^{-13}$ cm$^2$ s$^{-1}$.  This can be attributed to the effect of methane on ASW porosity upon deposition, since depositing methane together with water will slightly decrease the number of pores, slowing the diffusion. This is discussed in section 6.
 
 The $D(T)$ values found for methane diffusing on ASW grown at 50 K (experiments AA1-AA5) are about one order of magnitude smaller than those found for ASW ice grown at 30 K (exp. A5-A8). It is well known that, for background vapor deposition, the higher the growing temperature, the higher the average density of the ice obtained (\cite{Dohnalek2003}), i.e., the lower the porosity. Monte Carlo simulations  performed in this work corroborate this (see Figure \ref{fig:9}). Consequently, the experimental $D(T)$ values found indicate CH$_4$ diffuses slower through the pores of the more compact ASW ice.
 
\begin{figure}
    \centering
    \includegraphics[width=10cm]{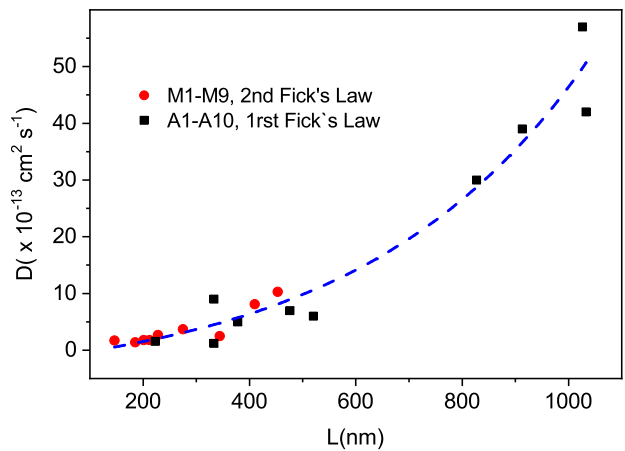}
    \caption{Diffusion coefficient of CH$_4$ at 50 K, for different ASW ice layer thicknesses. Red dots: Experiments M1-M9. Black starts: Experiments A1-A10. Dotted blue line: an exponential fit performed only to guide the eye.}
    \label{fig:6}
    \end{figure}

3) ASW ice layer thickness.
Although unexpected, a dependence of $D(T)$ on ASW layer thickness was observed in the experiments. All the diffusion coefficients obtained in this work for ASW grown at 30 K and T$_{iso}$= 50 K have been plotted in Figure \ref{fig:6} versus ASW layer thickness. It can be seen that below 600 nm, the diffusion coefficients fluctuations could be considered within experimental error. However, for layers above 600 nm, there is a clear increase in the diffusion coefficient with thickness. This behaviour has been observed by other authors, like Ghesquière and colaborators (\cite{Ghesquiere2015}) in analogous experiments performed to study CO$_2$ diffusion in water ice. Also May and coworkers (\cite{May2013}), investigating CH$_4$ diffusion through ASW layers grown on top, concluded that the CH$_4$ releasing mechanism was diffusive only up to a certain ASW layer thickness.
A possible reason for this behaviour could be the failure of the approximations considered in the Fick's law. In particular, to consider that $D(T)$ is constant along the $z$ direction within the whole ice layer thickness.

\subsection{CH$_4$ diffusion barrier}
\begin{figure} 
    \centering
    \includegraphics[width=10cm]{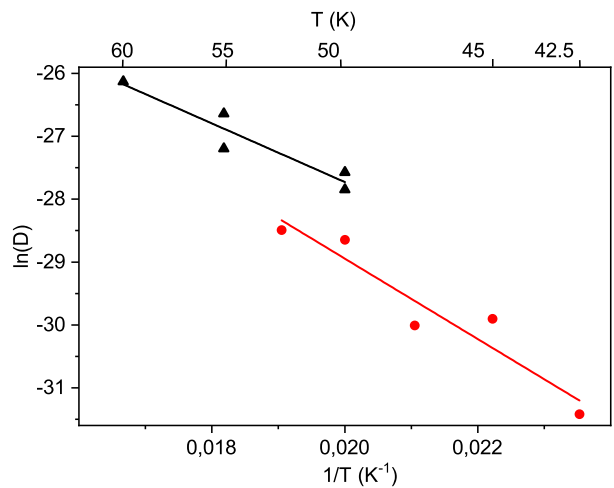}
    \caption{Arrhenius-type plot of the diffusion coefficients of CH$_4$ on ASW grown at 30 K (black triangles) and at 50 K (red dots).}
    \label{fig:7}
\end{figure}

Although only a few experiments were performed at temperatures different from 50 K, they could be used to constrain the diffusion barrier, assuming a single-barrier Arrhenius process and a $D(T)$ following Arrhenius equation:
\begin{equation}
D(T) = D_0 \exp\left(-{E_{d}/T}\right),
\label{eq:9}
\end{equation}
where $D_0$ is the pre-exponential factor, $E_{d}$ the diffusion energy barrier and $T$ the temperature. With the diffusion coefficients obtained for experiments M3, M4, M10, M11 and M12 at 50, 55 and 60 K, an Arrhenius-type plot has been performed to extract a diffusion barrier and pre-exponential factor. Also, experiments AA1-AA5, that give diffusion coefficients at temperatures between 42.5 and 52.5 K for ASW ice grown at 50 K,  have been fitted to the Arrhenius equation. Both fits are presented in Figure \ref{fig:7}. 

Energy barriers of 477 $\pm$ 91 K and 639 $\pm$ 118 K, and pre-exponential factors of ln($D_0$)= -18 $\pm$ 2 (log($D_0$)= -8.0  $ \pm$ 0.9) and ln($D_0$)= -16 $\pm$ 3 (log($D_0$)= -7.0  $ \pm$ 1.3) were found for both set of data, that correspond to ASW grown at 30 K and 50 K, respectively. 

In a previous work, He and coworkers (\citet{he2018}) measured the diffusion coefficient of CH$_4$ on ASW grown at 10 K and annealed to 70 K for 30 min. From their experiments, performed at temperatures between 17 K and 23 K, they found a E$_{d}$= 547 $\pm$ 10 K and ln(D$_0$)= -14.3 $\pm$ 0.5 (log(D$_0$)= -6.23  $ \pm $ 2). 
 
\begin{figure}
    \centering
    \includegraphics[width=10cm]{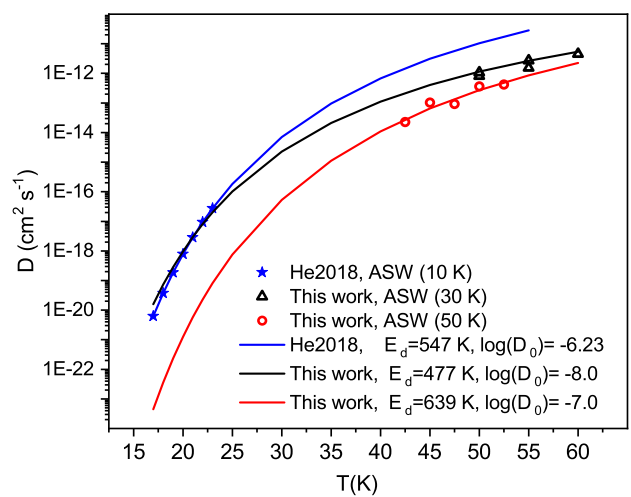}
    \caption{CH$_4$ diffusion coefficients obtained in this work compared with those in (\cite{he2018}). Scattered points: experimental data. The temperatures indicated in the legend refer to ASW generation temperature. Straight lines: fits extracted form Arrhenius plots.}
    \label{fig:8}
\end{figure}

Their data, compared with those obtained in this work, are presented with symbols in Figure \ref{fig:8}. The solid lines represent the extrapolations of the temperature dependence given by the Arrhenius equation. The differences between the three set of data could be due to differences in the morphology of the ASW ice, that were grown at 10 K, 30 K, or 50 K.

%---------------------------------------------
\section{Monte Carlo simulations}
We used a step-by-step Monte Carlo simulation to follow the formation of H$_2$O ices with CH$_4$ trapped gas through co-deposition. Our model is described in \cite{cazaux2015,cazaux2017}. H$_2$O and CH$_4$ molecules originating from the gas phase arrive at a random time and location on the substrate, and follow a random path within the ice. The arrival time depends on the rate at which gas species collide with the surface (section 5.1). The molecules arriving on the surface can be bound to the substrate and to other H$_2$O and CH$_4$ molecules through hydrogen bound and van der Waals interactions. In the present study, we used on-lattice KMC simulations. Other types of simulations, such as off-lattice KMC method have also been used to compute the porosity in ices \cite{Garrod2013}. While this method is more sophisticated than the present 
method since it allows to determine the distance of the species explicitly, we here consider the distance between water molecules to be equal, and concentrate on defining the binding energies as function of neighbours. Because our method does not 
compute the distance between molecules, the diffusion could be slightly different. However, we can provide diffusion barriers that can be compared with experimental results and used in theoretical models.

In the present study we consider two distinct cases on how to calculate the binding energies, depending on whether it concerns a water or a methane molecule. For water, the binding energy of each H$_2$O molecule depends linearly on its number of neighbours, as described in \cite{cazaux2015}. For methane, we consider that the binding energy of these molecules is already very high for one neighbor, and very close to the binding energies measured experimentally (Escribano et al., private communication). We therefore consider that the diffusion energy of methane, which is a fraction of the binding energy, is not depending on the number of neighbors, and use E$_{d} \sim 550~K$, as determined in \cite{he2018}. This value was chosen because we will compare our experimental results with those of \cite{he2018} (see section 6.3), despite different binding energies appear in the literature (see \cite{luna2014} and references therein). Depending on their diffusion energies, the H$_2$O and CH$_4$ molecules diffuse on the surface/in the ices. The diffusion is described in section 5.2. During warming-up and the waiting time, the molecules can evaporate from the substrate/ices. In order to reproduce experimental measurements, we deposit water and methane with a H$_2$O:CH$_4$ ratio of 10:1. We then increase the temperature after deposition with a ramp of 1~K every 10 seconds (which is 6~K/min, providing a ramp close to experimental values ranging between 5 to 10 ~K per minute). When the temperature of 50~K is reached, we let the system evolve with time at constant temperature and determine the number of molecules evaporating and staying in the ice. This allows a direct comparison with the measurements of section 3.

\subsection{Accretion}
In our model, we defined the surface as a grid with a size of 60$\times$60 sites. Low-density amorphous ice consists mainly of four coordinated tetrahedrally ordered water molecules. As already discussed in \cite{cazaux2015}, amorphous water ice is modeled using a grid in which the water molecules are organized as tetrahedrons, which implies that each water molecule has four neighbours. Molecules from the gas-phase arriving on the grid can be bound to the substrate. The accretion rate (in s$^{-1}$) depends on the density of the species, their velocity, and the cross section of  the surface, and can be written as:
\begin{equation}
R_{\rm{H_2O}} = n_{\rm{H_2O}} v_{\rm{H_2O}}  \sigma  \rm{S},
\end{equation}
\begin{equation}
R_{\rm{CH_4}} = n_{\rm{CH_4}} v_{\rm{CH_4}}  \sigma  \rm{S},
\end{equation}
where $v_{\rm{H_2O}}= \sqrt{\frac{8\ k\ T_{\rm{gas}}}{\pi\ m_{\rm{H_2O}}}} \sim 3.4 \times 10^4 \sqrt{\frac{T_{\rm{gas}}}{100}}$ \rm{cm~s}$^{-1}$ and $v_{\rm{CH_4}}=\sqrt{\frac{8\ k\ T_{\rm{gas}}}{\pi m_{\rm{CH_4}}}} \sim 3.6 \times 10^4 \sqrt{\frac{T_{\rm{gas}}}{100}}$ \rm{cm~s}$^{-1}$ are the thermal velocity of water and methane respectively, and $\sigma$, the cross section of the surface. S is the sticking coefficient that we consider to be unity in this study.  The distance between two sites is 1.58 \AA, but each water molecule occupies 1 site over 4 because of its four coordinates tetrahedral order (see \cite{cazaux2015}). The surface density of water molecules is what is typically assumed, i.e. $\sim (1/4 \times 1.58^{-2})  \AA^{-2} \sim$ 10$^{15}$ cm$^{-2}$. The cross section scales with the size of the grid considered in our calculations, which is 60$\times$60 sites, as $\sigma \sim (1.58 \ 10^{-8} \times 60)^2$ cm$^2$= 9 10$^{-13}$ cm$^2$. The deposition rate is therefore: $R_{\rm{acc}(H_2O)} = 3 \ 10^{-8}$ $n_{\rm{H_2O}}$ s$^{-1}$, and $R_{\rm{acc}(CH_4)} = 3.2 \ 10^{-8}$ $n_{\rm{CH_4}}$ s$^{-1}$, for T$_{\rm{gas}}$=100~K. In order to mimic experimental conditions with deposition rates of 1 nm/s $\sim$ 6 ML/s $\sim$ 450 molecules/s (we have 60$\times$60/8 molecules per ML), we set the density of H$_2$O molecules in the gas in cm$^{-3}$ as being $n_{\rm{H_2O}}$= 9 $\times$ 10$^{10}$ cm$^{-3}$. The density of CH$_4$ is scaled to the density of water to reproduce the experiments with H$_2$O:CH$_4$ 10:1.

\subsection{Diffusion}
We use a method to simulate diffusion similar than in \cite{cazaux2015}. The diffusion barrier is usually considered in models as a fraction of the binding energy. For water, the binding energy increases with the number of neighbors, and therefore the diffusion barrier also increases with the number of neighbors. For CH$_4$, on the other hand, we consider only one binding energy with water, and that the binding energy does not increase with the number of neighboring water molecules (consistent with the very small differences in binding energies seen in \cite{luna2014}). The diffusion rate for methane is therefore simply $R_{\rm{diff}} = \nu \exp\left({-\frac{E_{\rm{d}}}{\rm{T}}}\right)$, were $\nu$ is the pre-exponential factor and $E_d$ the diffusion barrier for methane.

For water, we compute the binding energy by adding the number of neighbors nn as E$_b$=nn $\times$ E$_b$(H$_2$O-H$_2$O), with E$_b$(H$_2$O-H$_2$O) $\sim$ 0.2 eV $\sim$ 2550~K \citep{brill1967,isaacs1999,dartois2013} with a pre-exponential factor of 10$^{13}$ s$^{-1}$ \cite{fraser2001}. We define the diffusion rates by calculating the initial energy of a molecule and its final energy in the possible sites where it can move. For a position (i,j,k) of a molecule in the grid, we calculate the associated binding energy E$_i$ and identify the possible sites where the molecule can diffuse to as i$\pm$1; j$\pm$1; k$\pm$1. The final binding energy $E_f$ is calculated as function of the neighbours present around this site. The diffusion rate, from an initial site with an energy $E_i$ to a final site with an energy $E_f$, is illustrated in \cite{cazaux2017}. The barrier to go from $E_i$ to $E_f$  is defined as follows if $E_i$ $\le$ $E_f$: 
\begin{equation}
E_d= \alpha \times \rm{min}(E_i,E_f), \hspace{1cm} if \  E_i < E_f
\end{equation}
If $E_i$ $>$ $E_f$, on the other hand, the barrier becomes:
\begin{equation}
E_d= \alpha \times \rm{min}(E_i,E_f) +\Delta E, \hspace{1cm} if \ E_i > E_f
\end{equation}
with $\Delta E$ = max($E_i$,$E_f$) - min($E_i$,$E_f$). By defining the barriers in such a manner, we do take into account microscopic reversibility in this study (\citealt{cuppen2013}), which implies that barriers to move from one site to another should be identical to the reverse barrier. The diffusion barriers scale with the binding energies with a parameter $\alpha$. For water, $\alpha$ sets the temperature at which water molecules can re-arrange and diffuse in the ices to form more dense ices. This parameter is found to be around 30~$\%$ for the water-on-water diffusion derived experimentally (\citealt{collings2003a}), and we obtained an alpha lower than 40~$\%$ in a previous study (\citealt{bossa2015}). For methane, \cite{he2018} derived an $\alpha$ ranging between 34 and 50~$\%$ and a binding energy ranging from 1100~K to 1600~K. In this study, we will use our simulations to constrain the diffusion barrier.

The diffusion rate for water, in s$^{-1}$, can be written as: 
\begin{equation}
R_{\rm{diff}} = 4 \times {\frac{\sqrt{\frac{E_f-E_s}{E_i-E_s}}}{\left(1+\sqrt{\frac{E_f-E_s}{E_i-E_s}}\right)^2}}\times \nu \exp\left({-\frac{E_{\rm{d}}}{\rm{T}}}\right), 
\end{equation}
where $\nu$ is the pre-exponential factor, T is the temperature of the substrate (water ice or CH$_4$ ice) and $E_s$ is the energy of the saddle point, which is $E_s$=(1-$\alpha$)$\times$ min($E_i$,$E_f$). This formula differs from typical thermal hopping because the energy of the initial and final sites are not identical (\citealt{cazaux2017}).

The pre-exponential factor is related with the vibrational frequency of a species in its site, and can be derived form \cite{Landau1966} expresion:
$ \nu=\sqrt{\frac{2\ N_s\ E_i}{\pi^2\times m_{CH4}}}$, where N$_s$ is the number of sites per surface area (10$^{15}$), m$_{CH_4}$ is the mass of methane and E$_i$ the binding energy. In this work we have taken $\nu$= 10$^{13}$ s$^{-1}$ for water and $\nu$=10$^{9}$ s$^{-1}$ for CH$_4$ (\cite{he2018}).

\subsection{Sublimation}
The molecules present on the surface can return into the gas phase because they sublimate. This desorption rate depends on the binding energy of the species with the surface/ice. As mentioned previously, the binding energy of a H$_2$O molecule depends on its number of neighbours, while we consider a unique binding energy for CH$_4$ independent of the number of neighbors. The binding energy $E_i$ of a molecule sets the sublimation rate as:
\begin{equation}
 R_{\rm{subl}}(X)=\nu \exp\left({-\frac{E_i}{T}}\right),
\end{equation}
where $\nu$ is the pre-exponential factor, which is taken as $\nu$=10$^{13}$ s$^{-1}$ for water, and $\nu$=10$^{9}$ s$^{-1}$ for CH$_4$ (\cite{he2018}). We used a pre-exponential factor similar than the one used for diffusion for simplicity. However, we also used the lowest binding energy of methane derived from experiments of 1100 ~K. In this sense, the desorption rate at 30~K is around R$_{\rm{subl}}(CH_4)= 10^{-7}$ s$^{-1}$, which is similar to a rate with a pre-exponential factor of 10$^{12}$ and a binding energy of 1300~K. We therefore use a sublimation rate in the range of what has been derived in other studies (\cite{he2018}). Using different pre-exponential factor and binding energy for sublimation do not change our results after deposition. Methane molecules desorb once they have diffused through the ice to reach the surface, and therefore our results depend on the diffusion rate rather than sublimation rate.  

\section{Theoretical results}
\subsection{Water ice morphology and porosity}
\begin{figure*}
    \centering
    \includegraphics[width=6cm]{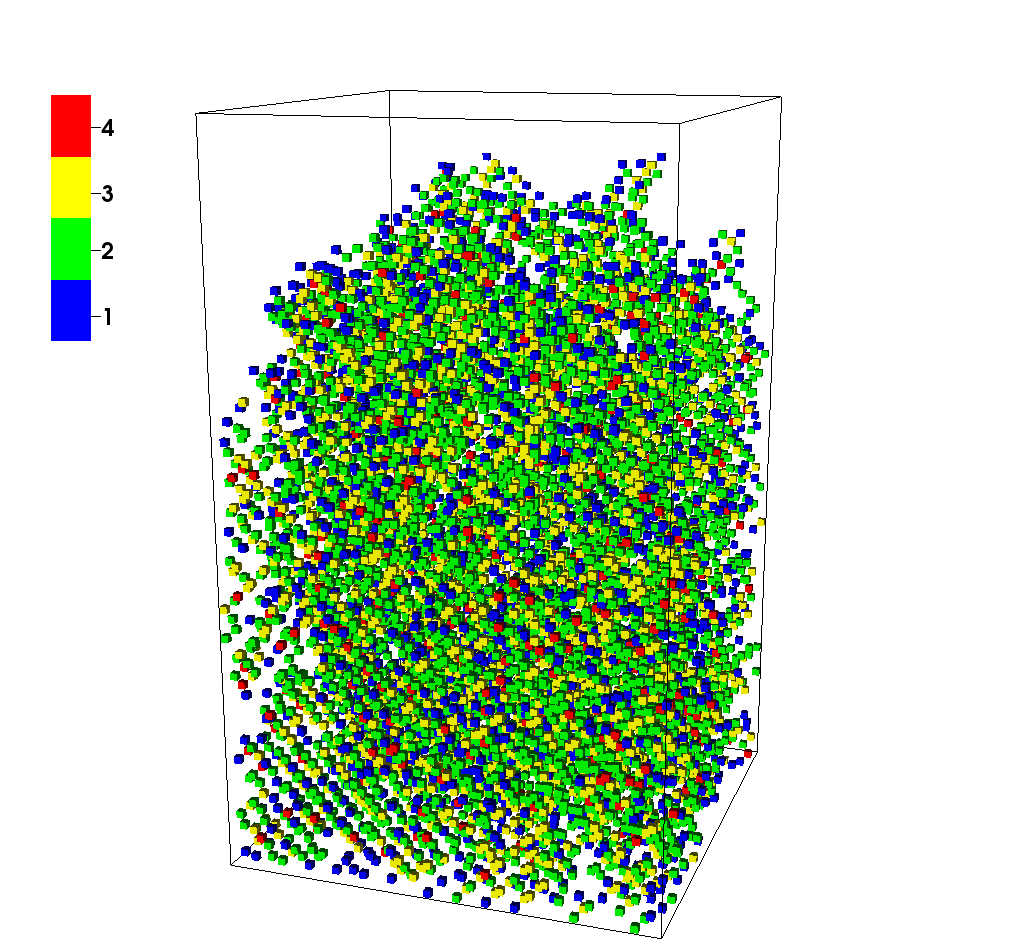}
    \includegraphics[width=6cm]{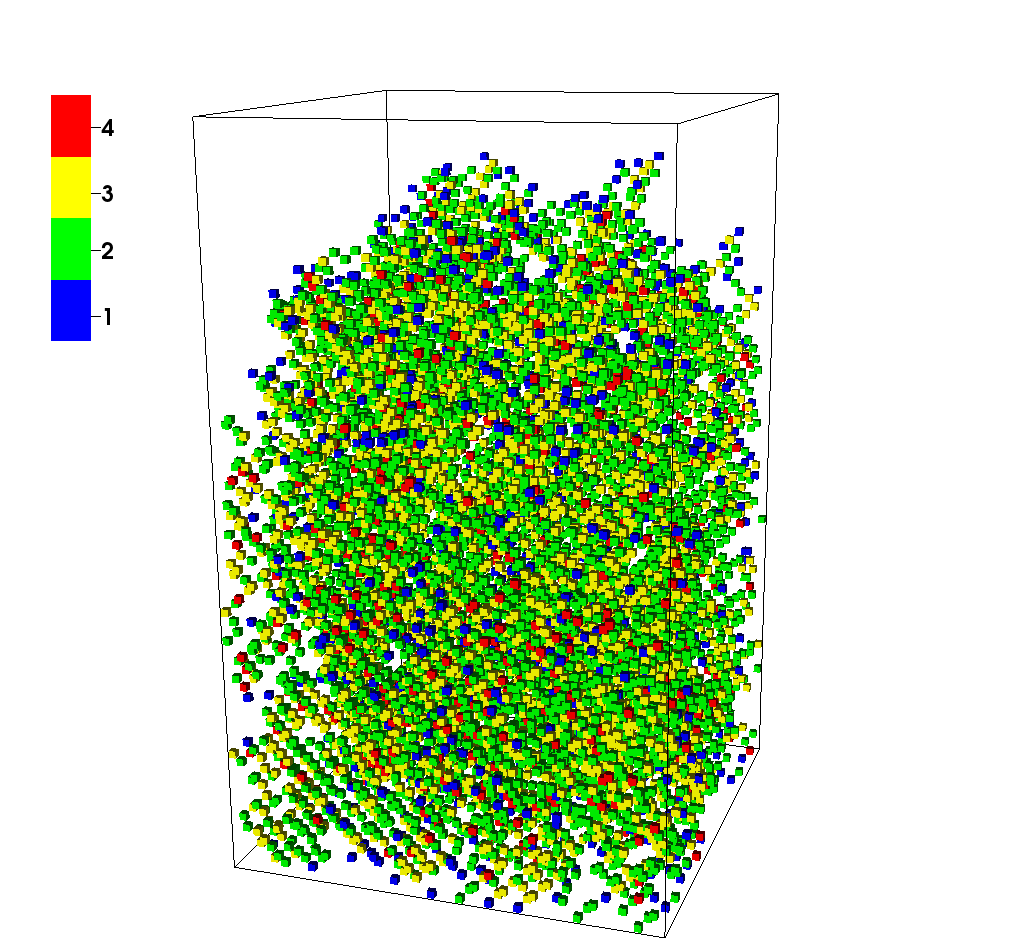}    \includegraphics[width=6cm]{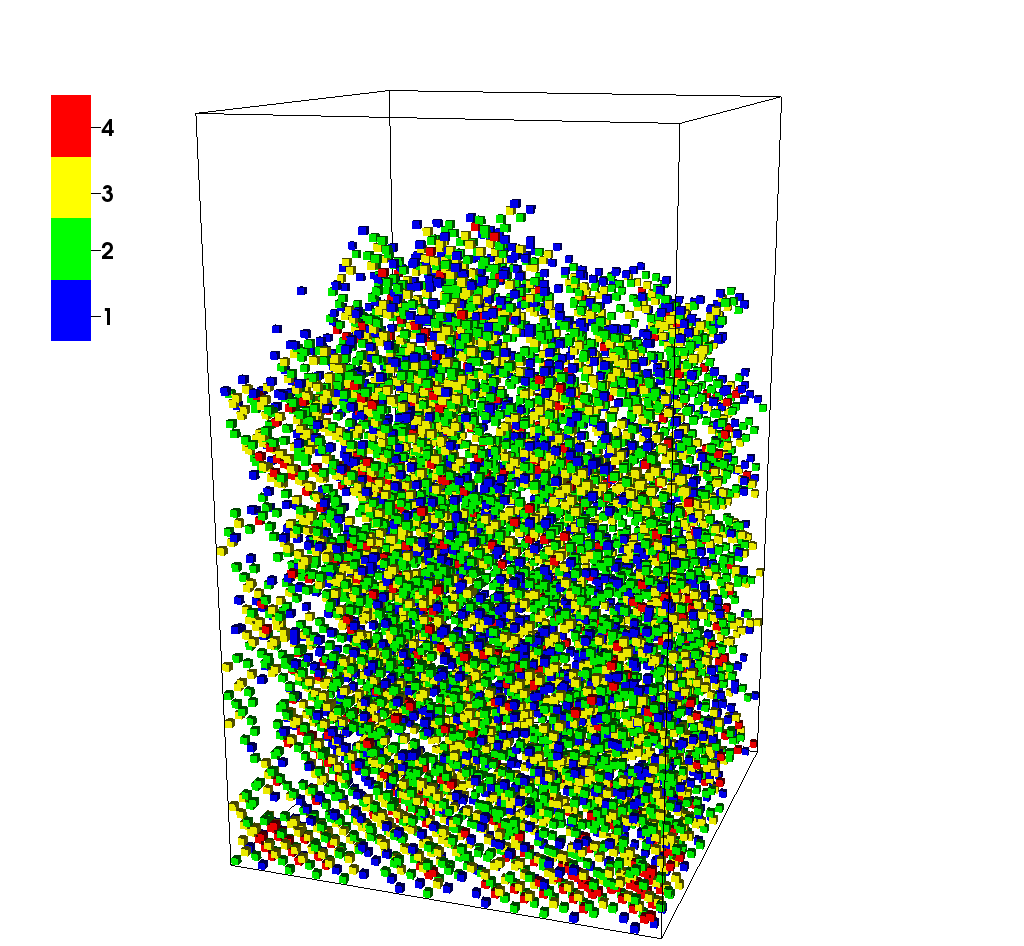}
    \includegraphics[width=6cm]{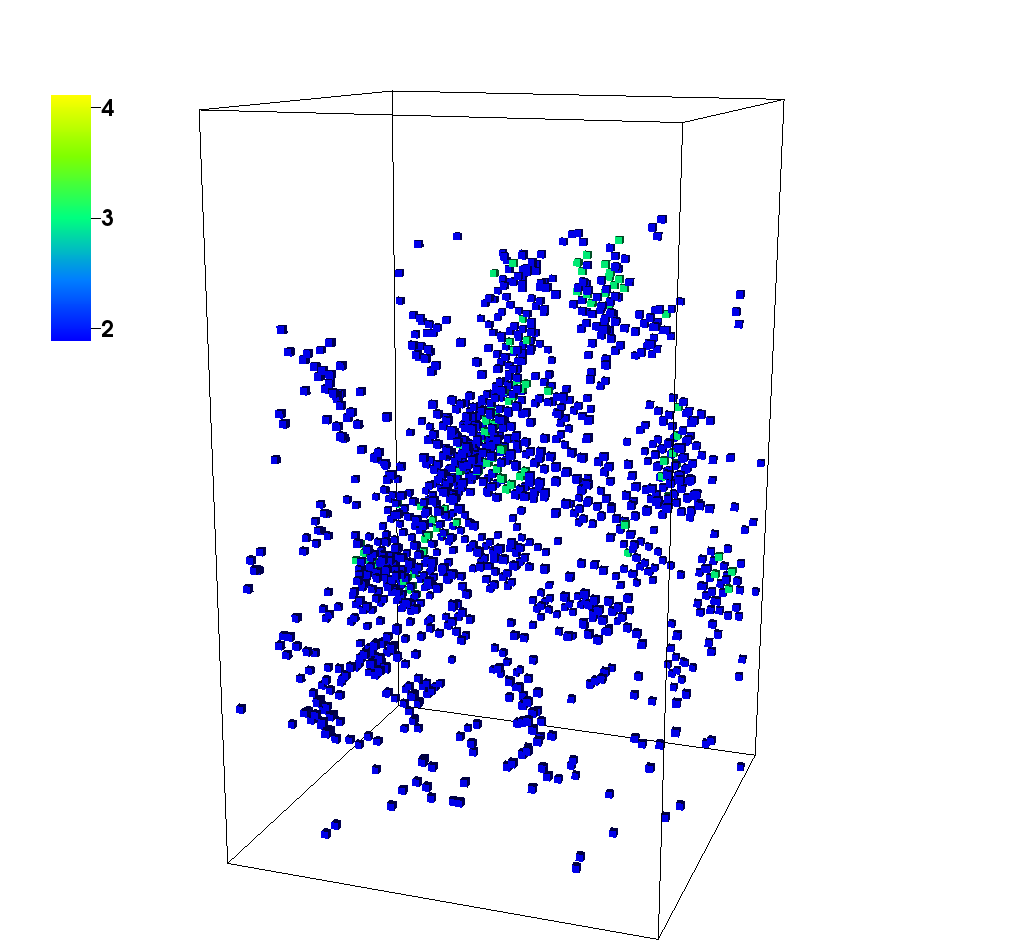}
    \includegraphics[width=6cm]{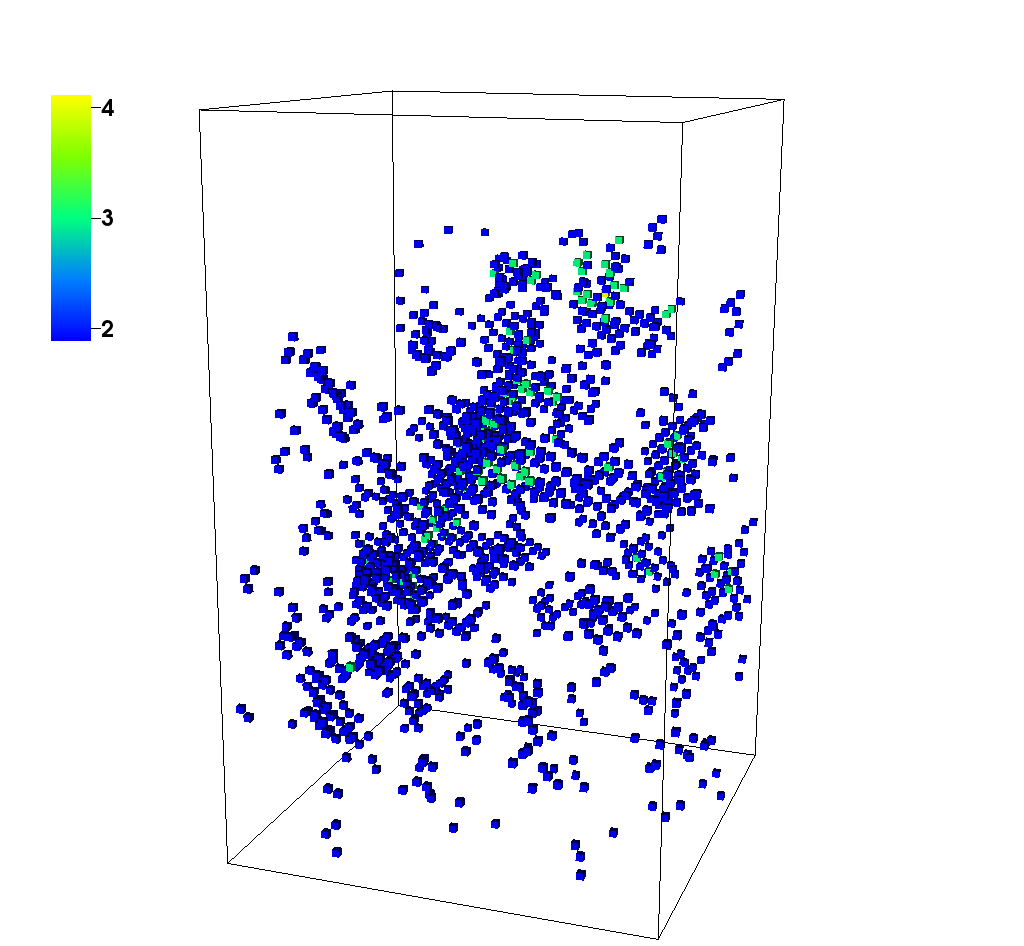}
    \includegraphics[width=6cm]{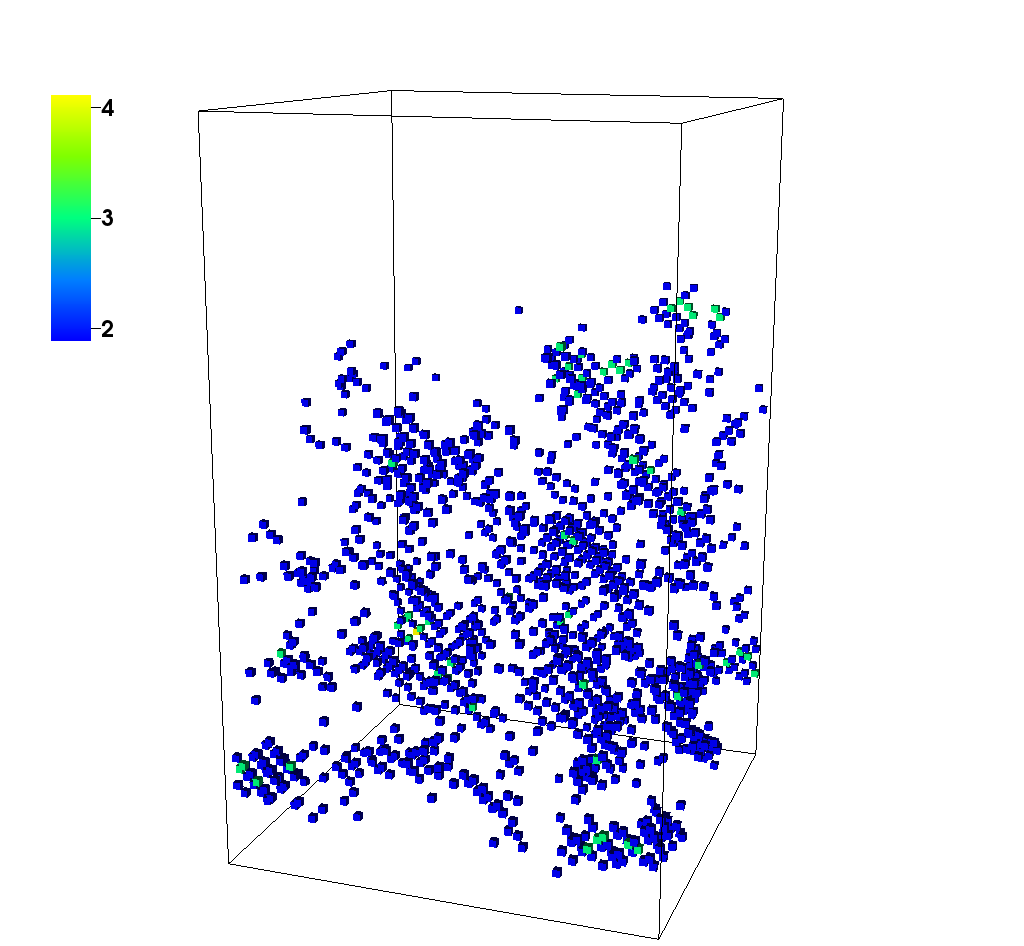}
    \caption{Water morphology and porosity under our experimental conditions. Top panels show water ice being deposited at 30~K (left panel), and then being heated to 50~K (middle panel) and water ice being deposited at 50~K (right panel). Each square represent a water molecule and the color represent the number of neighbour for each molecule. Blue squares represent water molecules with 1 neighbour, green with 2 neighbours, yellow with 3 and red with 4 neighbors. The bottom panels show the size of the pores in the water ice, which is the negative image of the top panel (showing the emptiness in ice). The color shows the number of grid cells empty around the cell (corresponding to 1.6~\AA). The figures show deposition at 30~K (left panel), after being heated to 50~K (middle panel) and being deposited at 50~K (right panel). }
    \label{fig:9}
\end{figure*}

The experiments have been performed for a deposition of water and methane at 30~K with a rate of around 1 nm/s and a ratio of H$_2$O:CH$_4$ of 10:1. The ice was then heated until 50~K and the diminution of the IR absorbance spectra has given the amount of CH$_4$ desorbing from the ice. Measuring the concentration of CH$_4$ with time can directly give a constraint on the diffusion coefficient, as shown in the previous sections. CH$_4$ diffuses through the pores within the water ice, which implies that the porosity of water ice could play a role in the diffusion. To grasp how water ice changes within the conditions explored in the present experiments, we show the morphology of water ice with our simulations when deposited at 30~K (Figure \ref{fig:9}, top left panel) and heated at 50~K (Figure \ref{fig:9}, top middle panel), and being deposited at 50~K (Figure \ref{fig:9}, top right panel). Each colored square represents a water molecule, and the color indicates the number of neighbours for each of them, which defined the biding energy. Blue corresponds to 1 neighbor (binding energy of $\sim$2500~K), yellow to 2 neighbors (5000~K), green 3 neighbors ($\sim$7500~K) and red to 4 neighbors ($\sim$10000~K). At 30~K, the water ice present mostly water molecules with 1 and 2 neighbors as the figure \ref{fig:9} (top left) shows mostly blue and green colors. However, when the ice is heated to 50~K, the color in the figure \ref{fig:9} (top middle) changes to green and yellow (less blue appear) showing that water molecules did rearrange and molecules with 1 neighbor moved to another location with more neighbors. If the deposition occurs at 50~K (top right), then the dominating colors are blue and green as in the first figure. This is because this simulation shows the state of the ice just after accretion and molecules are still re-organising. The bottom panels show the size of the pores in the water ice. These panels are the negative images of top panels and show the emptiness within the ices. The color shows the size of the pores, with a minimum of 2 (blue) to 4 (yellow). A pore size of 2 indicates that 2 grid cell around the pore are empty, while 3 and 4, indicates larger pores of 3 or 4 grid cell sizes (in radius). One grid cell represent $\sim$1.6 \AA, and therefore pores vary from 3.2 to 6.4 \AA. The different panels show the pores network after deposition at 30~K (left panel), after the ice deposited at 30~K being heated to 50~K (middle panel) and for ice being deposited at 50~K (right panel). The pores in the case of deposition at 30~K, are less and less connected than if the water is subsequently heated to 50~K. The evolution of the pores network show how diffusion within the pores will take place when methane is added into the water ice. When water ice is deposited at 50~K, the pores are smaller and less connected, which indicates that diffusion would be less efficient.

\subsection{Methane diffusion in thin water ice: effect of diffusion barrier and porosity.}
\begin{figure*}
    \centering
    \includegraphics[width=8cm]{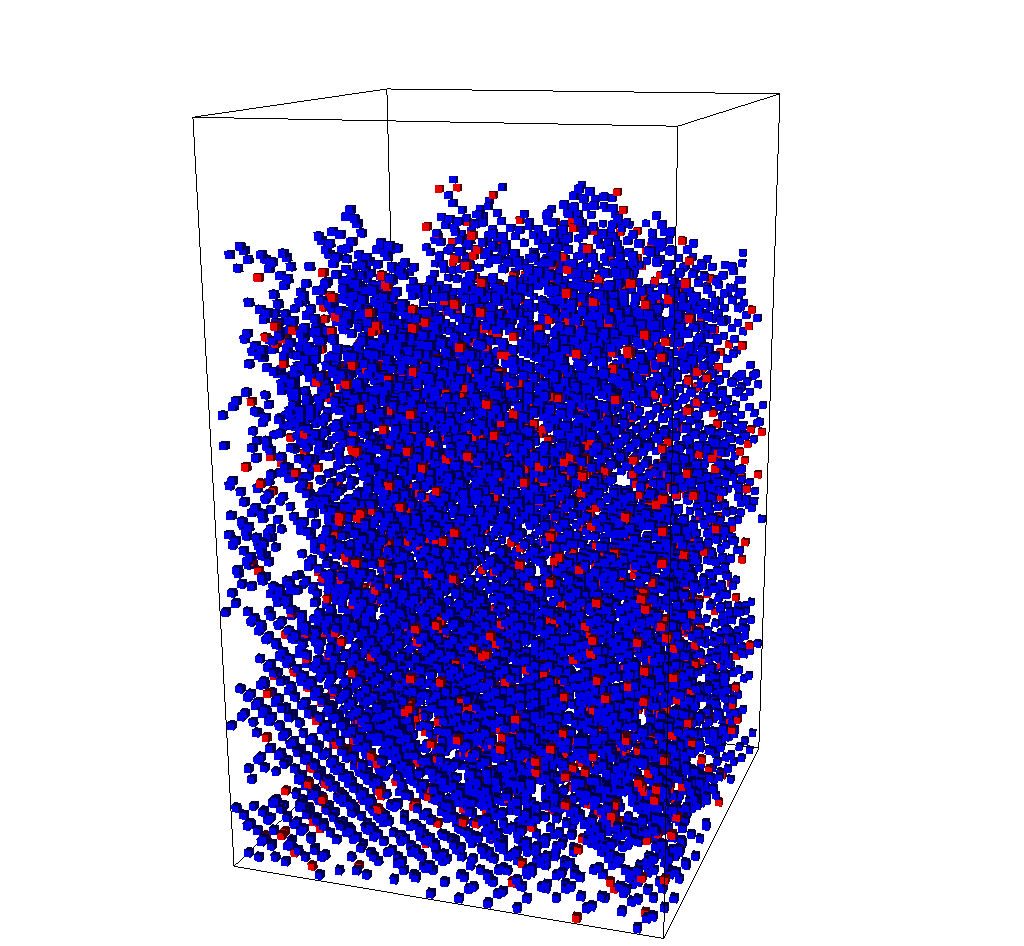}
    \includegraphics[width=8cm]{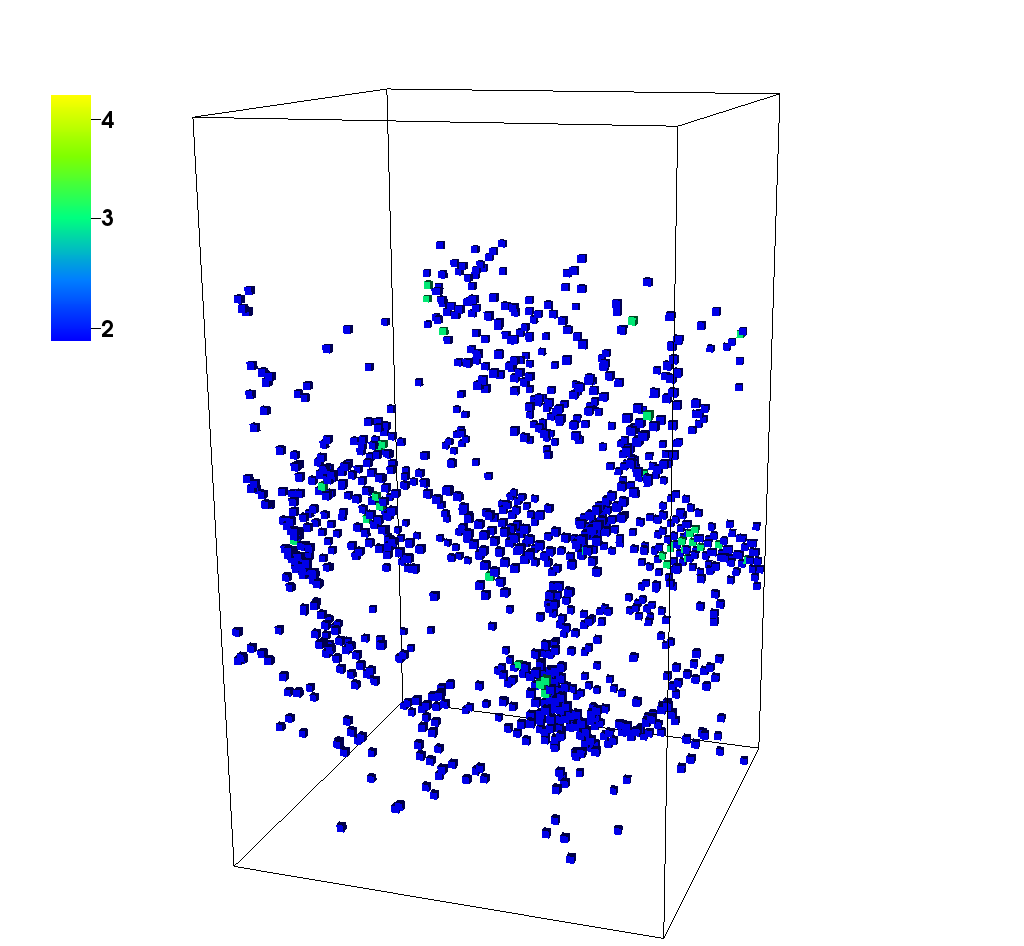}
    \caption{Water ice (blue) with methane (red) after deposition at 30~K (left panel). The right panel shows the size of the pores in the water ice mixed with methane. Pore sizes vary from $\sim$ 6 to 10 \AA\ in diameter.}
    \label{fig:10}
\end{figure*}

In order to understand the trapping and diffusion of methane in water ices, we performed Monte Carlo simulations for thin ices deposited at 30~K, heated until 50~K and with a waiting time similar to those observed experimentally. Methane is deposited simultaneously with the water, with a ratio of H$_2$O:CH$_4$ of 10:1. In this section we use thin ices of 10 nm thick to study the effect of the diffusion barriers and of the water ice morphology. In Figure \ref{fig:10} we illustrate how methane is mixed within water ice in one of our simulations. In the left panel, the blue boxes represent the water molecules, while the red boxes show the presence of methane molecules. The ice mixture is represented just after deposition at 30~K. In this simulations, we use a diffusion barrier of E$_{d}$ = 0.4 E$_i$ for water, and a diffusion barrier of E$_{d}$ = 660~K for methane (note that for water E$_i$ depends on the number of neighbors, while the diffusion barrier for methane is always the same). The thickness of the ice is slightly larger when methane is included in the simulations. This is due to the fact that at 30~K methane diffuses as much as water with one neighbor, which implies that the reorganisation is faster than if the ice was composed of 100$\%$ water ice. The porosity, presented in Figure \ref{fig:10}, right panel, seems to be less important than in the case of pure water (left bottom panel in Figure \ref{fig:9}). This could be due to the fact that the molecules do diffuse slightly more than for 100~\% water, which decrease the coalescence of pores to form networks. It is possible to relate this with the diffusion coefficients found for sequential and codeposited ices, and commented in section 4. In sequential ices, CH$_4$ will be diffusing through a pure AWS layer, slightly more porous than the ASW generated by codepositon, and therefore a faster CH$_4$ diffusion is expected.

\begin{figure*}
    \centering
    \includegraphics[width=6cm,angle=-90]{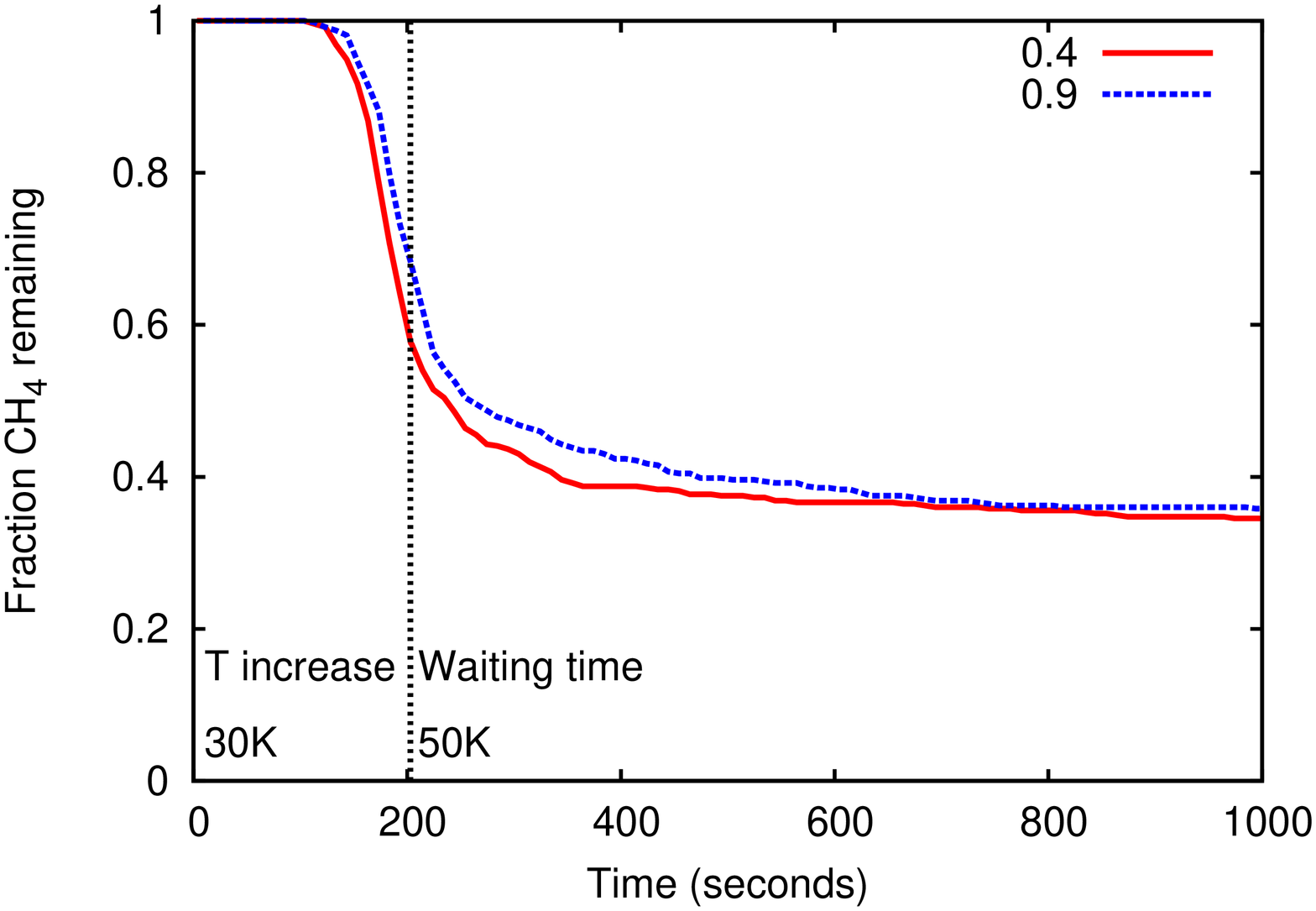}
    \includegraphics[width=6cm,angle=-90]{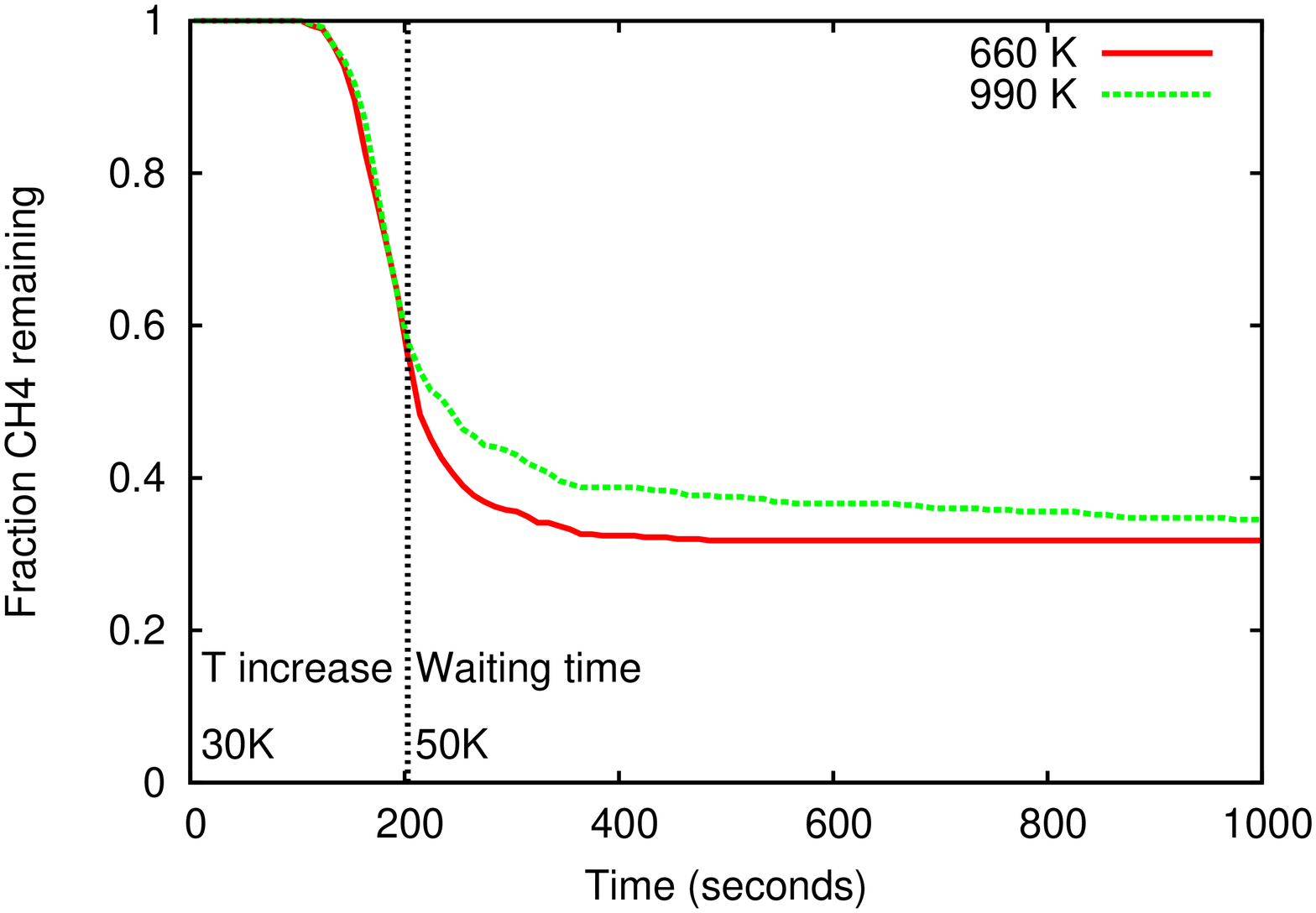}
    \caption{Left panel: effect of water diffusion on the amount of methane remaining in the ice. In red E$_d$(H$_2$O)=0.4 E$_i$ while in blue E$_d$(H$_2$O)=0.9 E$_i$. Right panel: effect of methane diffusion on the amount of methane remaining in the ice. In red E$_d$(CH$_4$)=660 K while in green  E$_d$(CH$_4$)=990 K}.
    \label{fig:11}
\end{figure*}

The effect of the reorganization of water is illustrated in Figure \ref{fig:11}, left panel. In this figure we fix the diffusion of methane with E$_{d}$=990~K. The diffusion of water is set to two different values $\alpha$(H$_2$O)=0.4 (red) and 0.9 (blue) which gives a diffusion barrier of  E$_{d}$=$\alpha$(H$_2$O) E$_i$. This figure shows that the reorganization of water has an effect on the diffusion of methane. This is due to the fact that methane diffuses in pores up to the surface of the ice. The reorganization of water affects the presence of pores and the way they are connected. A slower diffusion of water will create less pores, and consequently, the diffusion of methane will be somewhat slower. The diffusion of methane is presented in Figure \ref{fig:11}, right panel, for E$_{d}$=660~K (red) and 990~K (green). The diffusion of water is set to $\alpha$(H$_2$O)=0.4 which gives  E$_{d}$=0.4 E$_i$. A lower mobility of methane (E$_{d}$=990~K) implies that the diffusion is slower and that the number of methane desorbing from the ices is smaller. This is seen in the figure when comparing to a faster diffusion given by a smaller barrier of E$_{d}$=660~K, as shown in red. Therefore, both the mobility of water and methane are key parameters to constrain the diffusion coefficient of methane in water ices.\\

\subsection{Thick ice: constraining the diffusion from experimental data.}
\begin{figure}
%\begin{figure*}
    \centering
    \includegraphics[width=6cm,angle=-90]{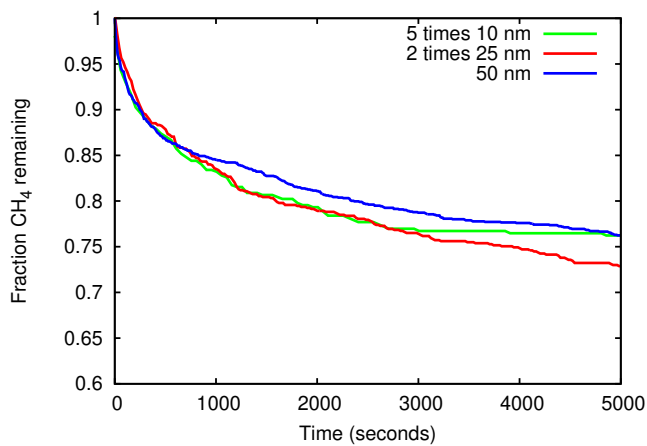}
    \caption{Comparison of the fraction of CH$_4$ remaining for a 50 nm (in blue) compared with the fraction in 2 blocks of 25 nm attached (red) and 5 block attached (green).}
    \label{fig:12}
\end{figure}

Most of the experimental measurements used to determine the diffusion have been obtained on ices with thickness larger than 180 nm. Our simulations can be made for thin ices, or for thick ices when diffusion is slow, but cannot be performed for fast diffusion in thick ices due to large computing times. In the present section, we show an alternative solution to simulate the amount of CH$_4$ remaining in thick ices. We made calculations for ices of 10 nm thick and used several of these thin ices to scale the results to a thick ice. Since we can compute the number of CH$_4$ evaporating from a 10 nm ice, we can compute the amount of CH$_4$ evaporating from several blocks of thin ices attached to each other to mimic a thick ice. If the fraction of CH$_4$ evaporating from the 10 nm ice thick is NEV$_{CH_4}$, then the fraction of CH$_4$ evaporating from two blocks of 10 nm thick ice is NEV$_{CH_4}$+ NEV$_{CH_4}^2$ (the second term shows the fraction reaching the top block from the bottom block, and then crossing the top block to evaporate). If we would have $n$ blocks of ice attached to each other, we could estimate the fraction of CH$_4$ evaporating as NEV$_{CH_4}$+NEV$_{CH_4}^2$+NEV$_{CH_4}^3$+..+NEV$_{CH_4}^n$. In order to validate our method, we compute the amount of CH$_4$ staying in the ice with one block of 10 nm, one of 25 nm and one of 50 nm. The amount of CH$_4$ remaining in the ice is computed for 2 $\times$  25 nm as 1- 1/2 $\times$ (NEV$_{CH_4}$+NEV$_{CH_4}^2$), with NEV$_{CH_4}$ is the number of CH$_4$ evaporating from a block of 25 nm thick, while for 5$\times$10 nm as 1- 1/5 $\times$ (NEV$_{CH_4}$+NEV$_{CH_4}^2$+NEV$_{CH_4}^3$+NEV$_{CH_4}^4$+NEV$_{CH_4}^5$), with NEV$_{CH_4}$ is the number of CH$_4$ evaporating from a block of 10 nm thick. The simulations for a 50 nm thick ice are shown in figure \ref{fig:12} in blue, while the simulations for 2$\times$25 nm are shown in red, and 5$\times$10 nm, in green. We can see that this method shows very good agreement to estimate the amount of CH$_4$ remaining, with a very small difference on short timescales below 1000 seconds, and a difference of a few percent for larger timescales. We can therefore use this method to reproduce the amount of CH$_4$ remaining in thick ices in the experiments. These simulations were made for slow diffusion where E$_{d}$ = 990~K for methane and E$_{d}$ = 0.9 E$_i$ for water.

\begin{figure*}
    \centering
    \includegraphics[width=6cm,angle=-90]{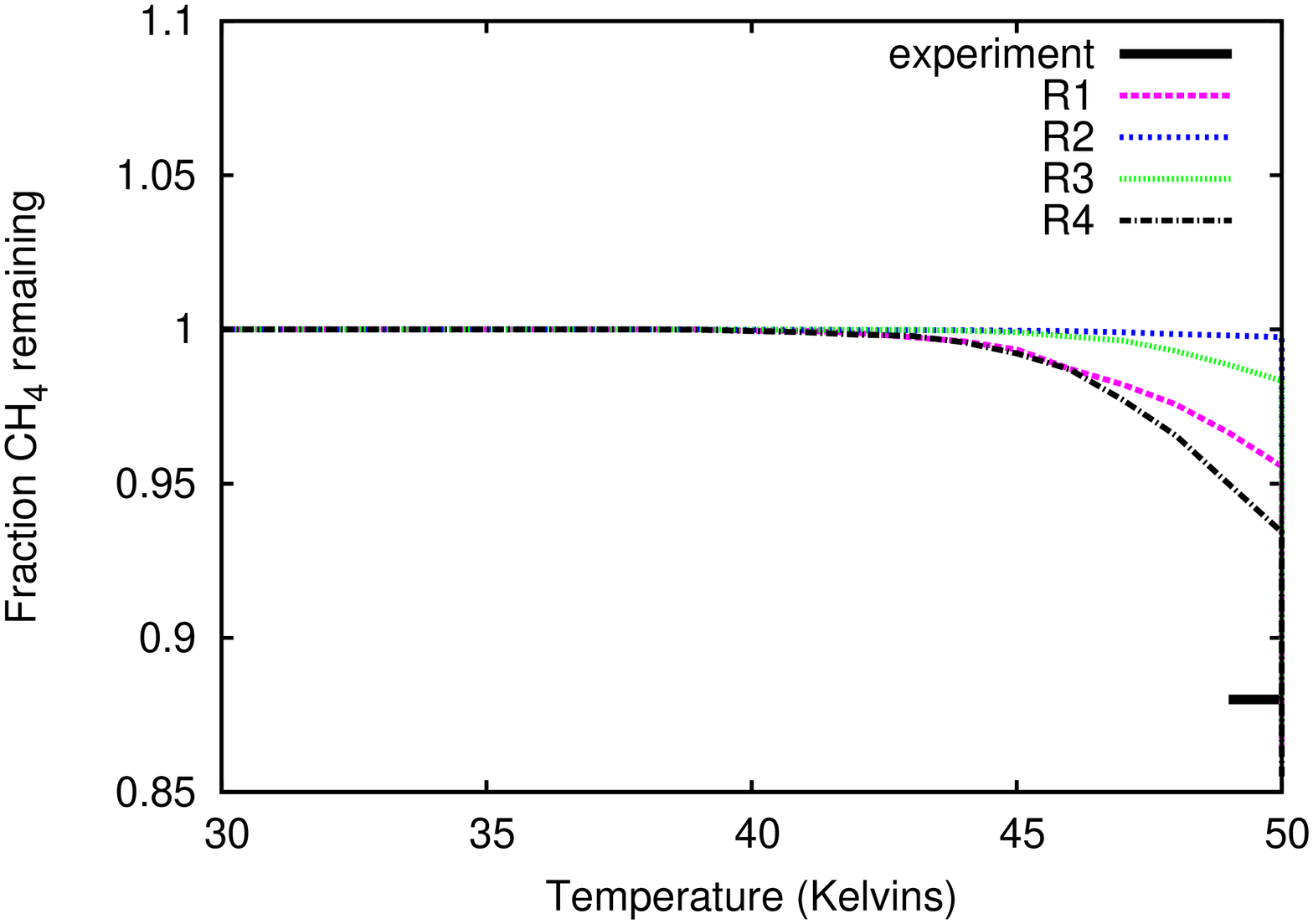}
    \includegraphics[width=6cm,angle=-90]{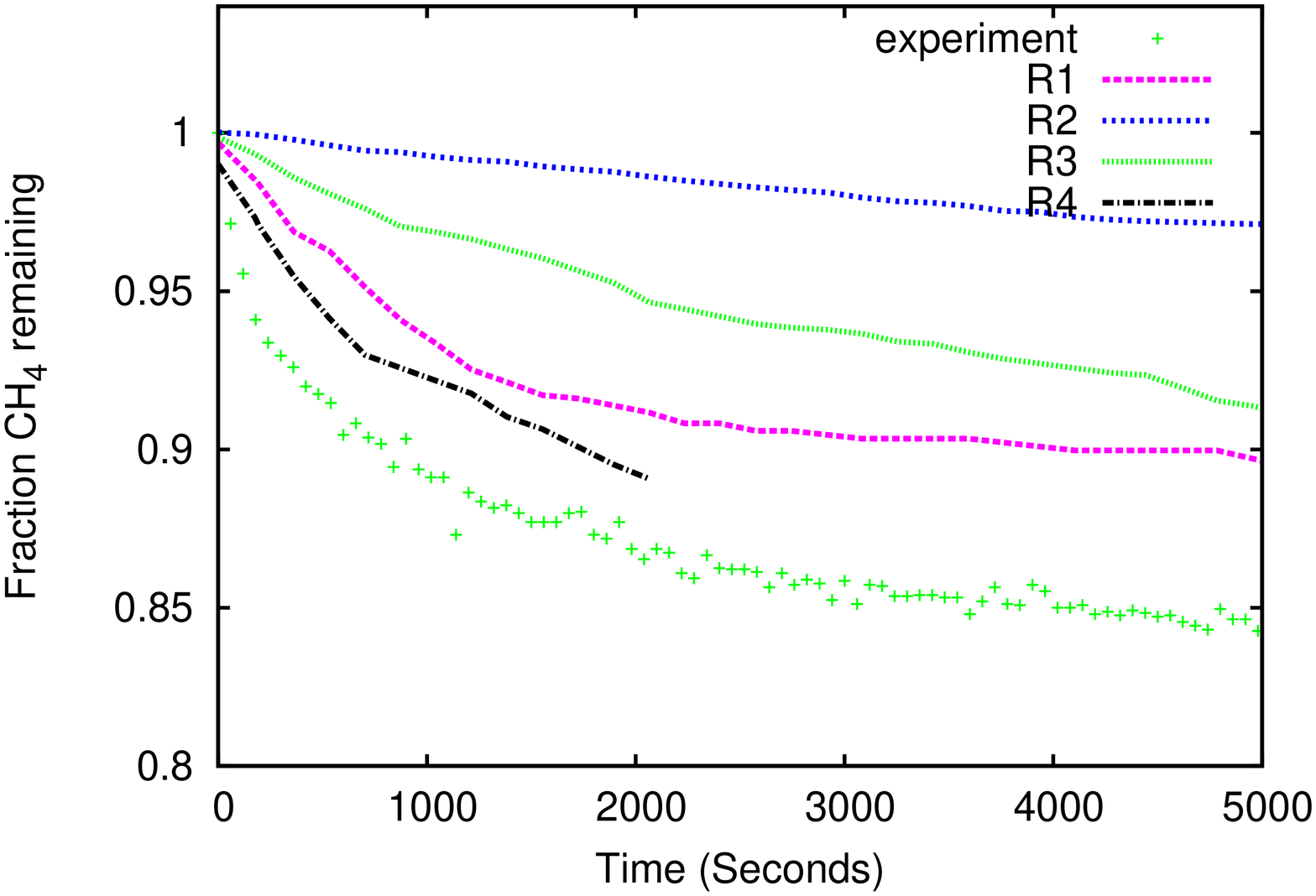}
    \caption{Left panel: fraction of CH$_4$ remaining in the ices during the TPD. The experiments show that 13.8 $\%$ of CH$_4$ was present initially at 30~K while 12.1 $\%$ remained at 50~K. Right panel: fraction of CH$_4$ remaining in the ices during the waiting time after reaching 50~K.The green points show the experimental values.}
    \label{fig:13}
\end{figure*}

\begin{figure}
    \centering
    \includegraphics[width=9.5cm]{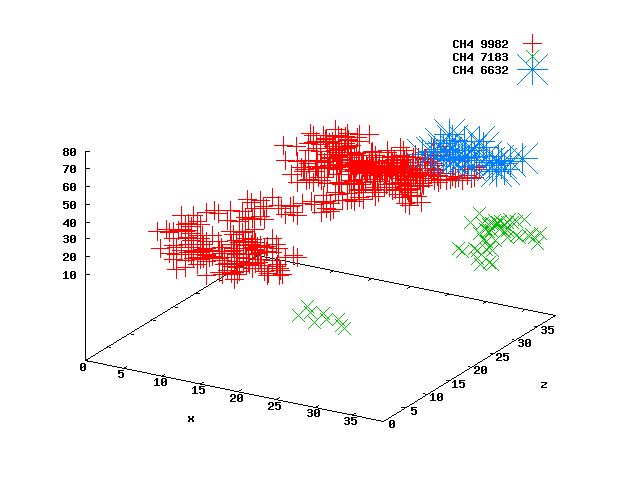}
    \caption{Selection of three methane molecules moving in the grid during the entire experiment. The path clearly illustrates a preferred mobility in the pores. The fact that the green CH$_4$ moves form a side of the box to the other side is due to side effects.}
    \label{fig:14}
\end{figure}

%\end{figure*}
In the simulations presented in this section, we aim to reproduce the experimental results from section 3 in order to constrain the diffusion coefficient of methane on ASW deposited at 30~K. We use in our simulations the results obtained in the experiment M7, and use different diffusion rates from \cite{he2018} (mentioned as R1) and the ones derived in this work. We converted the D$_0$ obtained previously with the formula: D$_0$=$\frac{\nu a^2}{4}$, as shown in \cite{he2018}, where $\nu$ is the pre-exponential factor, and a=3\AA\ is the distance between two sites. We used the rate derived by \cite{he2018} of R1=2 10$^9 \exp{\frac{-547}{T}}$ s$^{-1}$, and derived the rates from the present study, R2=4 10$^7 \exp{\frac{-477}{T}}$ s$^{-1}$ and R3=4 10$^8 \exp{\frac{-639}{T}}$s$^{-1}$. We also consider another rate R4, in order to show the effect of the increase of the pre-exponential factor on the fraction of CH$_4$ trapped. In this case, R4=2 10$^9 \exp{\frac{-477}{T}}$ s$^{-1}$. Figure \ref{fig:13}, left panel, shows the fraction of CH$_4$ remaining after heating, using the rates mentioned above. In the experiments, 0.88 $\%$ of methane is still present in the ice when reaching 50~K (thick red stripe). Our model cannot account for such a loss with the rates considered. In order to reproduce the measurements, a larger rate (larger $\nu$ or lower activation energy) should be used. This implies that in the present case, the diffusion is hindered because many molecules are diffusing back and forth in pores before being able to arrive to the surface and desorb. In other words, the microscopic diffusion of methane is many times faster than the diffusion measured experimentally. Indeed, methane molecules are constantly diffusing within the ices, which can add up to a very long path while the actual distance travelled is orders of magnitude lower. This is illustrated by Figure \ref{fig:14}, which shows the movement of three methane molecules taken randomly in the ices. The three methane molecules move within the pores without reaching the surface, and the total distance they explored in the ice is very different from the actual path they follow. The symbols show the position of each molecule (each molecule represented by a colored cross) during the entire simulation. These are the results from simulations using $\nu$=2  10$^9$ s$^{-1}$ and E$_{d}$=660~K, for a 10 nm thick ice. The simulations are following the movement of individual molecules during the heating ramp and waiting time, which lasts in total $\sim$ 1000 seconds. Note that we consider that molecules can move from one side of the box to another (if a molecule reaches the limit of the box and moves outside, its position is placed to the other side of the box. This happens with the green molecule). Figure \ref{fig:13}, right panel, shows our Monte Carlo results compared to the experimental measurements for a 185 nm thick ice. The simulations are overestimating the fraction of CH$_4$ remaining in the ice for the four rates considered. Note that for R4, the calculation is computationally expensive, we only show the results of the simulations until 2000 seconds. This implies that to reproduce the experimental results, the diffusion should be faster (larger pre-exponential factors or lower activation energy) than the diffusion obtained using the Fick's law.  Our results show that the rates extracted from the present study as well as the rate from \cite{he2018} do not reproduce the experimental data with our Monte Carlo simulations.

\subsection{Diffusion versus porosity}
\begin{figure}
    \centering
    \includegraphics[width=6cm,angle=-90]{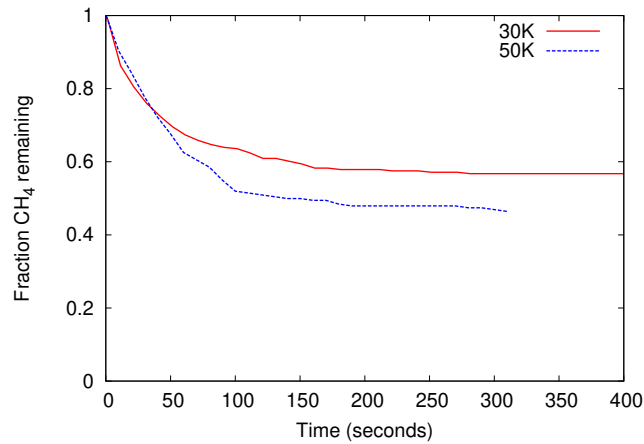}
    \caption{Fraction of methane remaining in the ice after 50 K when deposited at 30~K and at 50~K.}
    \label{fig:15}
\end{figure}

In this section we investigate the effect of porosity on the diffusion of methane in the ice. We deposited ices at 30~K and 50~K in our simulations, and determined the fraction of methane in the ices as a function of time after 50~K. These results are presented in Figure \ref{fig:15}. For the ices deposited at 30~K, we increased the temperature until 50~K, and then determine the amount of CH$_4$ with waiting time (red line), while for the ices deposited at 50~K, we determine the amount of CH$_4$ after the deposition is completed (green line). In these simulations, the mobility for water ice is set as E$_{d}$=0.4 E$_i$ for water and E$_{d}$=990~K for methane. Note that on both simulations the diffusion rates are the same and only the deposition temperature changes. Our results show that the porosity changes the diffusion of methane within the ice. A more porous medium, such as the one created when deposition occurs at 30~K, allows for a faster diffusion and therefore less CH$_4$ remain in the ice. On the other hand, a less porous medium, such as the one created when deposition occurs at 50~K, slows down the diffusion and consequently more CH$_4$ molecules remain in the ice. This result is in agreement with experimental findings shown in Figure \ref{fig:8}, where the diffusion coefficient obtained for ices grown at 50~K is smaller (slower diffusion) than the diffusion of ices grown at 30~K. This shows that the porosity plays a role in the diffusion of molecules within the ices, larger and connected pores favoring a higher mobility.

\section{Conclusions}

We have investigated the surface diffusion of methane on amorphous solid water, and how it is affected by the porosity of the ASW structure. 

Methane diffusion coefficients have been measured to vary between 10$^{-13}$ and 10$^{-12}$ cm$^2$ s$^{-1}$, for temperatures between 42 K and 60 K. It was observed that different ASW structures modify the CH$_4$ diffusion coefficient up to one order of magnitude.

Experiments and simulations show periods of time where the rate of sublimation of methane is constant. This implies that the methane concentration gradient remains constant for these time intervals, and the first Fick's law can be used. 

Monte Carlo simulations of simultaneous and sequential H$_2$O:CH$_4$ ices at 30 K, show that a more compact ASW structure is formed in codeposited experiments. The diffusion coefficients measured for CH$_4$ trapped in codeposited ASW are smaller (slower diffusion) than those found in sequential experiments. Therefore, it is possible to say that when a CH$_4$ reservoir is diffusing through a pure ASW layer formed on top, the diffusion will be faster than in homogeneously mixed  H$_2$O:CH$_4$ ices.

Measured diffusion coefficients indicate that CH$_4$ diffuse faster on ices grown at 30 K than in ices grown at 50 K. Monte Carlo simulations show that ices grown at 50 K present smaller and less interconnected pores than ices grown at 30 K. Therefore, it can be concluded that larger and more interconnected pores favour methane diffusion.

We used Monte Carlo simulations in order to better understand the experimental results,  and to estimate the effect of porosity on the diffusion rates. We showed the diffusion rates derived experimentally using Fick's laws do not reproduce the experimental results with our Monte Carlo simulations, considering the range of diffusion values within the experimental uncertainty. The experimental results can be reproduced only using a diffusion rate at least 50 times higher, which would imply for the present study a pre-exponential factor ranging between 10$^{10}$-10$^{11}$ s$^{-1}$. They are consistent with factors derived from \cite{Landau1966} expression (see section 5.2.).
%the expression used to derive the pre-exponential factor from \cite{Landau1966}: $ \nu=\sqrt{\frac{2\ N_s\ E_i}{\pi^2\times m_{CH4}}}$,} where N$_s$ is the number of sites per surface area (10$^{15}$), m$_{CH_4}$ is the mass of methane and E$_i$ the binding energy. 
Our work therefore shows that the diffusion coefficient obtained from the Fick's law is a diffusion at the macroscopic level. This can be used to determine the overall loss of methane in ices on large timescales (for example when modelling ices of comets, or moons). However, when modelling reactivity or ice evolution in microscopic level with kinetic models (Monte Carlo simulations, rate equations, etc.) the diffusion used should be much higher and the pre-exponential factor can be derived as in \cite{Landau1966}.
%The discrepancy between macroscopic (Fick's law) and microscopic view of diffusion comes from the fact that microscopic models account for back and forth diffusion,in the medium while macroscopic see the effective distance that the molecules travelled.    
The discrepancy between macroscopic (Fick's law) and microscopic view of diffusion comes from the fact that microscopic models account for back and forth diffusion, counting single diffusion hops in all directions in the medium. In contrast, macroscopic diffusion sees the effective distance that methane molecules travelled, describing CH4 as a continuum medium diffusing due to a concentration gradient.

In summary, our work shows that the microscopic diffusion of methane is many times faster than the macroscopic diffusion measured experimentally.

\begin{acknowledgements}
Funds from the Spanish MINECO/FEDER FIS2016-77726-
C3-1-P and C3-3-P projects are acknowledged.   
\end{acknowledgements}


\bibliographystyle{aa} % style aa.bst
\bibliography{CH4_DIFFUSION} % your references Yourfile.bib


\end{document}